\documentclass[%
 reprint,
 amsmath,amssymb,
 aps,
superscriptaddress,
]{revtex4-1}

\usepackage{graphicx}
\usepackage{dcolumn}
\usepackage{bm}
\usepackage{tikz}
\usepackage[bottom]{footmisc}
\usetikzlibrary{positioning, calc, arrows, shapes}
\usepackage[colorlinks=true, linkcolor=blue, citecolor=black, urlcolor=blue]{hyperref}

\newcommand{\trento}{{\tt T\raisebox{-.5ex}{R}ENTo }}
\newcommand{\PHSD}{{\rm \tt PHSD} }
\newcommand{\Duke}{{\rm \tt Duke} }
\newcommand{\Nantes}{{\rm \tt Nantes} }
\newcommand{\CpQCD}{{\rm \tt Catania-pQCD} }
\newcommand{\CQPM}{{\rm \tt Catania-QPM} }
\newcommand{\LBT}{{\rm \tt CCNU-LBT} }

\definecolor{phsd}{RGB}{255,0,0}
\definecolor{catania}{RGB}{51,153,255}
\definecolor{nantes}{RGB}{204, 0, 204}
\definecolor{lbt}{RGB}{255, 153, 51}
\definecolor{duke}{RGB}{0,153,0}
\definecolor{comment}{RGB}{255,153,255}

\begin{document}

\title{Cracking the difference of estimating heavy quark transport coefficients in a Quark-Gluon Plasma}

\author{Yingru Xu}
\email{yx59@phy.duke.edu}
\affiliation{Department of Physics, Duke University, Durham, North Carolina 27708, USA}

\author{Steffen A. Bass}
\affiliation{Department of Physics, Duke University, Durham, North Carolina 27708, USA}

\author{Pierre Moreau}
\affiliation{Institute for Theoretical Physics, Johann Wolfgang
	Goethe Universit\"{a}t, D-Frankfurt am Main, Germany}

\author{Taesoo Song}
\affiliation{Institut f\"ur Theoretische Physik, Universit\"at Gie\ss{}en, Heinrich-Buff-Ring 16, D-35392 Gie\ss{}en, Germany}

\author{Marlene Nahrgang}
\affiliation{SUBATECH UMR 6457 (IMT Atlantique, Universit\'{e} de Nantes,
	IN2P3/CNRS), 4 Rue Alfred Kastler, F-44307 Nantes, France}

\author{Elena Bratkovskaya}
\affiliation{Institute for Theoretical Physics, Johann Wolfgang Goethe Universit\"{a}t, D-Frankfurt am Main, Germany}
\affiliation{GSI Helmholtzzentrum f\"{u}r Schwerionenforschung GmbH, Planckstrasse 1, D-64291 Darmstadt, Germany}

\author{Pol Gossiaux}
\affiliation{SUBATECH UMR 6457 (IMT Atlantique, Universit\'{e} de Nantes,
	IN2P3/CNRS), 4 Rue Alfred Kastler, F-44307 Nantes, France}

\author{Jorg Aichelin}
\affiliation{SUBATECH UMR 6457 (IMT Atlantique, Universit\'{e} de Nantes,
	IN2P3/CNRS), 4 Rue Alfred Kastler, F-44307 Nantes, France}
\affiliation{FIAS, Johann Wolfgang Goethe Universit\"{a}t, D-Frankfurt am Main, Germany}

\author{Shanshan Cao}
\affiliation{Department of Physics and Astronomy, Wayne State University, Detroit, MI 48201, USA}

\author{Vincenzo Greco}
\affiliation{Department of Physics and Astronomy, University of Catania, Via S. Sofia 64, I-95125 Catania, IT}
\affiliation{INFN-LNS, Laboratori Nazionali del Sud, Via S. Sofia 62, I-95125 Catania, IT}

\author{Gabriele Coci}
\affiliation{Department of Physics and Astronomy, University of Catania, Via S. Sofia 64, I-95125 Catania, IT}
\affiliation{INFN-LNS, Laboratori Nazionali del Sud, Via S. Sofia 62, I-95125 Catania, IT}

\author{Klaus Werner}
\affiliation{SUBATECH UMR 6457 (IMT Atlantique, Universit\'{e} de Nantes,
	IN2P3/CNRS), 4 Rue Alfred Kastler, F-44307 Nantes, France}

\date{\today}

\begin{abstract}
Heavy flavor observables provide valuable information on the properties of the hot and dense Quark-Gluon Plasma (QGP) created in ultra-relativistic nucleus-nucleus collisions. Various microscopic models have successfully described many of the observables associated with its formation. Their transport coefficients differ, however, due to different assumptions about the underlying interaction of the heavy quarks with the plasma constituents, different initial geometries and formation times, different hadronization processes and a different time evolution of the QGP. In this study we present the transport coefficients of these models and investigate systematically how some of these assumptions influence the heavy quark properties at the end of the QGP expansion. For this purpose we impose on these models the same initial condition and the same model for the QGP expansion and show that both have considerable influence on $R_{AA}$ and $v_2$. 
\end{abstract}

\maketitle

\section{\label{sec:introduction} Introduction}
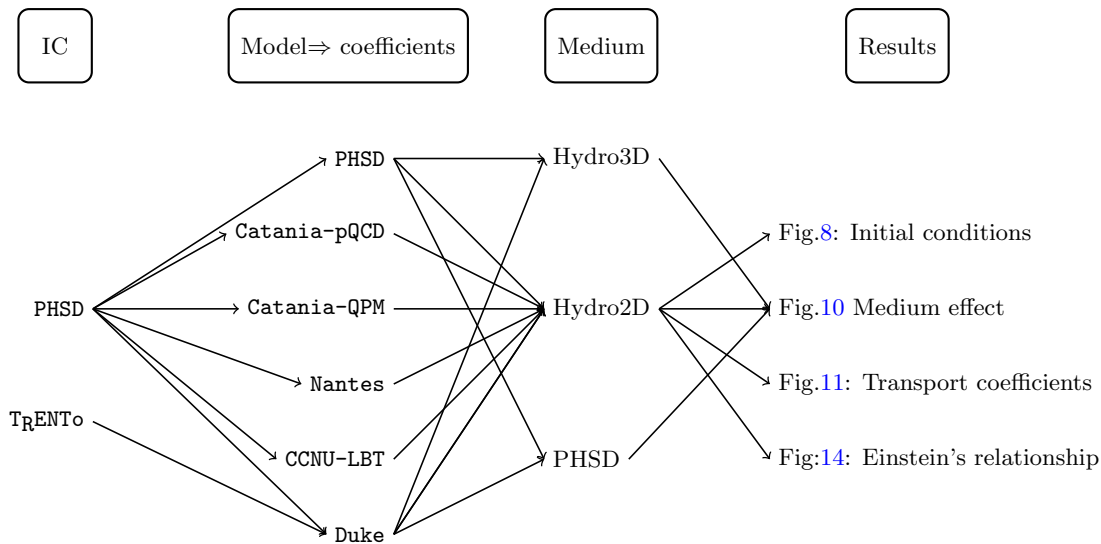
\begin{figure*}
	\begin{tikzpicture}
	\tikzset{
		mynode/.style={rounded corners, draw=black, top color=white, bottom color=white, thick, inner sep=0.5em, minimum size=3em, text centered},
		boxBlue/.style={rounded corners, align=center,inner sep=1ex,fill=dblue!20,text=black!30,text=dblue},
		boxRed/.style={rounded corners, align=center,inner sep=1ex,fill=dred!20,text=black!30,text=dred},
	}
	
	\node [mynode, anchor=middle, left=5cm of current page.center, yshift=1.5cm] (IC) {IC};
	\node [anchor=middle, left=5cm of current page.center, yshift=-2cm] (IC-PHSD) {\PHSD};
	\node [anchor=middle, left=5cm of current page.center, yshift=-3.5cm] (IC-trento) {\trento};

	\node [mynode, anchor=middle, left=0cm of current page.center, yshift=1.5cm] (model) {Model$\Rightarrow$ coefficients};
	\node [anchor=middle, left=1cm of current page.center, yshift=0cm] (PHSD) {\PHSD};
	\node [anchor=middle, left=1cm of current page.center, yshift=-1cm] (Catania-p) {\CpQCD};
	\node [anchor=middle, left=1cm of current page.center, yshift=-2cm] (Catania-Q) {\CQPM};
	\node [anchor=middle, left=1cm of current page.center, yshift=-3cm] (Nantes) {\Nantes};
	\node [anchor=middle, left=1cm of current page.center, yshift=-4cm] (LBT) {\LBT};
	\node [anchor=middle, left=1cm of current page.center, yshift=-5cm] (Duke) {\Duke};

	\node [mynode, anchor=middle, right=1cm of current page.center, yshift=1.5cm] (medium) {Medium};
	\node [anchor=middle, right=1cm of current page.center, yshift=0cm] (Hydro3D) {Hydro3D};
	\node [anchor=middle, right=1cm of current page.center, yshift=-2cm] (Hydro2D) {Hydro2D};
	\node [anchor=middle, right=1cm of current page.center, yshift=-4cm] (PHSDMedium) {PHSD};


	\node [mynode, anchor=middle, right=5cm of current page.center, yshift=1.5cm] (medium) {Results};
	\node [anchor=middle, right=4cm of current page.center, yshift=-1cm] (r1) {Fig.\ref{fig:HQ_v2_IC}: Initial conditions};
	\node [anchor=middle, right=4cm of current page.center, yshift=-2cm] (r3) {Fig.\ref{fig:3medium_b6fm} Medium effect};
	\node [anchor=middle, right=4cm of current page.center, yshift=-3cm] (r2) {Fig.\ref{fig:Raa_v2_6fm}: Transport coefficients};

	\node [anchor=middle, right=4cm of current page.center, yshift=-4cm] (r4) {Fig:\ref{fig:Raa_v2_6fm_ER}: Einstein's relationship};
	
	\draw[->, line width=0.6pt, black] (IC-PHSD.east) to (Duke.west);
	\draw[->, line width=0.6pt, black] (IC-trento.east) to (Duke.west);
	\draw[->, line width=0.6pt, black] (Duke.east) to (Hydro2D.west);

	\draw[->, line width=0.6pt, black] (IC-PHSD.east) to (PHSD.west);
	\draw[->, line width=0.6pt, black] (IC-PHSD.east) to (Catania-p.west);
	\draw[->, line width=0.6pt, black] (IC-PHSD.east) to (Catania-Q.west);
	\draw[->, line width=0.6pt, black] (IC-PHSD.east) to (Nantes.west);
	\draw[->, line width=0.6pt, black] (IC-PHSD.east) to (LBT.west);

	\draw[->, line width=0.6pt, black] (Duke.east) to (Hydro2D.west);
	\draw[->, line width=0.6pt, black] (Nantes.east) to (Hydro2D.west);
	\draw[->, line width=0.6pt, black] (LBT.east) to (Hydro2D.west);
	\draw[->, line width=0.6pt, black] (Catania-p.east) to (Hydro2D.west);
	\draw[->, line width=0.6pt, black] (Catania-Q.east) to (Hydro2D.west);
	\draw[->, line width=0.6pt, black] (PHSD.east) to (Hydro2D.west);

	\draw[->, line width=0.6pt, black] (Duke.east) to (PHSDMedium.west);
	\draw[->, line width=0.6pt, black] (PHSD.east) to (PHSDMedium.west);

	\draw[->, line width=0.6pt, black] (Duke.east) to (Hydro3D.west);
	\draw[->, line width=0.6pt, black] (PHSD.east) to (Hydro3D.west);

	\draw[->, line width=0.6pt, black] (Hydro2D.east) to (r3.west);
	\draw[->, line width=0.6pt, black] (Hydro3D.east) to (r3.west);
	\draw[->, line width=0.6pt, black] (PHSDMedium.east) to (r3.west);
	
	\draw[->, line width=0.6pt, black] (Hydro2D.east) to (r1.west);
	\draw[->, line width=0.6pt, black] (Hydro2D.east) to (r2.west);
	\draw[->, line width=0.6pt, black] (Hydro2D.east) to (r4.west);
	\end{tikzpicture}
	\caption{A skeleton showing each ingredients that needs to be taken into consideration during the implementation of heavy quark evolution, which could affect the estimation of the heavy quark transport coefficients in the QGP phase.}
	\label{fig:different_component}
\end{figure*}

One of the major efforts of heavy-ion physics aims at creating a phase of deconfined quarks and gluons (the Quark-Gluon plasma -- QGP) and estimating the characteristic transport properties of the QGP~\cite{Akiba:2015jwa}. 
Due to its short lifetime, estimation of the QGP properties relies on the comparison between the experimental data and  theoretical calculations which implement the interactions inside the medium.

Heavy quarks are among the most important probes for the study of the QGP medium \cite{Rapp:2009my,Andronic:2015wma}. 
They are primarily produced in the early stage of the heavy ion collisions via hard QCD scattering processes, and the production cross section can be calculated using a pertubative QCD approach.
During their propagation through the medium, heavy quarks interact with the medium and lose energy. 
Various approaches have been developed to describe the interaction between the heavy quarks and the surrounding medium.

It is useful to characterize this
 interaction by  a few transport coefficients: the drag coefficients $\eta_D$, the spatial diffusion coefficient $D_s$, the momentum transport coefficients $\kappa_L, \kappa_T, \hat{q}$~\cite{Banerjee:2011ra, Zakharov:2000iz,Gyulassy:2002yv,Baier:1996sk,Wang:2001ifa}, etc. The reduction of the interaction to a few transport coefficients has two advantages. On the one side, for each approach that models the interaction between heavy quarks and the medium, by comparing the calculation for different choices of transport coefficients with the experimental data, one should be able to constrain the values and functional form of the interaction strength.
 On the other side, it allows for a comparison among various approaches which have been advanced to describe the heavy quark - QGP interaction.

A comparison of these calculations in order to understand the different outcomes so far has been ambiguous \cite{Andronic:2015wma,Prino:2016cni}. This is not only due to the relatively large uncertainties in the experimental measurements -- which will be improved in the near future -- but also,  more essentially,  due to the interplay between different assumptions when one models the full sequential evolution of heavy quarks in heavy-ion collisions: initial conditions, pre-equilibrium dynamics~\cite{Chesler:2013urd}, formation time, time evolution of the QGP, in-medium propagation, hadronization~\cite{Braaten:2002yt, Lin:2003jy}, hadronic final state interactions~\cite{Laine:2011is,He:2011yi}  (Fig.~\ref{fig:different_component}). 
Each of these requires sophisticated modeling and introduces assumptions that need to be justified, as it may compensate differences in the description of the elementary heavy quark - QGP interaction.
It is therefore rather challenging to disentangle the contributions from different stages of the evolution apart from those of the heavy quark medium interactions, and truthfully summarizes the theoretical uncertainties regarding the determination of the transport coefficients in the QGP phase.
Despite all the difficulties, much effort has been made during the past to compare among different theoretical calculations and investigate the deviation, such as a systematic comparison of different charm quark transport coefficients in a static medium contributed by the JET-HQ collaboration~\cite{Cao:2018ews}, a broad investigation on the heavy quark evolution modeling components conducted by the EMMI framework~\cite{Rapp:2018qla}.

In this work we would continue the investigation by controlling variables and quantifying how differing model assumptions other than the heavy quark medium interactions contribute to the observed variability in the extracted heavy quark transport coefficients. We evaluate the response of the charm quark evolution inside a realistic QGP medium using different sets of transport coefficients -- which are estimated by multiple microscopic transport models -- in a standard Langevin evolution framework.
Fig.~\ref{fig:different_component} provides a schematic overview of how we separate each ingredient out and investigate its respective impact. The six sets of transport coefficients analyzed in this study are estimated from the following microscopic transport models:
\begin{itemize}
	
	\item \PHSD~\cite{Song:2015sfa,Song:2007fn}: the Parton-Hadron-String Dynamics transport approach,
based on off-shell Kadanoff-Baym equations (in first order gradient expansion). Heavy quarks interact with the off-shell quasi-particles whose masses and widths are evaluated to reproduce the lattice QCD EoS. Heavy quarks scatter with light quarks and gluons elastically, with the running coupling being determined by the scale of the temperature.
	
	\item \CpQCD~\cite{Plumari:2012ep,Das:2015ana}: Full space-time transport model for describing both heavy quark and massless light quark and gluon evolution based on the relativistic Boltzmann equation which is solved numerically by means of the test-particle method.
	Heavy quarks interact elastically with the bulk constituents where the scattering cross section is calculated at leading-order pQCD with a temperature-dependent running coupling $\alpha_s$ and Debye screening mass $m_D$.

	\item \CQPM~\cite{Scardina:2017ipo, Plumari:2011mk}: The evolution of heavy quarks and bulk partons is described by means of a Boltzmann equation similarly to \CpQCD. In this case, in order to account for non perturbative interactions, light quarks and gluons are dressed with thermal masses according to a quasi-particle prescription and the T-dependence of $\alpha_s$ is tuned to match the lattice QCD EoS.

	\item \Nantes~\cite{Gossiaux:2008jv,Nahrgang:2013saa}: a pQCD inspired running $\alpha_s$ Monte Carlo at Heavy Quark approach {\tt MC@sHQ}, where heavy quarks interact with the medium constituents (thermal massless partons) according to their scattering rate, using a linearized Boltzmann equation. The running coupling is implemented as reaching saturation at small $Q^2$ momentum transfer and the matrix elements are simplified by adopting an effective scalar propagator $\frac{1}{t - \kappa \tilde{m}_D^2(T)}$, with the Debye mass $\tilde{m}_D(T)$ evaluated self-consistently. Both collisional and collisional+radiative energy loss versions are implemented in the {\tt MC@sHQ} model. In this study, the collisional energy loss only version is used.

	\item \LBT~\cite{Cao:2016gvr,Cao:2017hhk}: Linearized Boltzmann dynamics of heavy quarks inside a hydrodynamical model describing the QGP medium. Heavy quarks interact with the medium constituents (thermal massless partons) according to pQCD scattering rates, where the running coupling is dependent on the momentum transfer scale. The gluon radiation rate utilizes the higher-twist formula, which is the same as in the \Duke model. 
	
	\item \Duke~\cite{Cao:2013ita,Cao:2015hia,Xu:2017hgt}: Improved Langevin dynamics of heavy quark inside a QGP medium modeled by fluid dynamics, incorporating both collisional and radiative energy loss. No specific assumption regarding the nature of the medium degrees of freedom is made, as the medium is defined by local temperature and flow velocity and is simulated by a hydrodynamical model. The interaction between heavy quarks and the medium is characterized by diffusion coefficients, which follow an empirical parametrization and are determined by Bayesian inference of the experimental measurements.
\end{itemize}
Note that we do not intend to perform a comparison of the different microscopic interaction mechanisms, which requires a more sophisticated study and will be conducted in the future.

The paper is structured as follows: in Sec.~\ref{sec:model} we will briefly review the Langevin dynamics that is used as a reference evolution model and present the transport coefficients extracted from the different models under discussion. 
Section~\ref{sec:in-medium} will investigate the effects that different modeling ingredients have on the outcome of the calculations. A summary will be addressed in Sec.~\ref{sec:summary}.

\section{\label{sec:model} Model description}
\subsection{\label{subsec:Langevin} Langevin dynamics}
We investigate the contribution from different components by inserting the extracted transport coefficients into a standard Langevin approach~\cite{He:2013zua}:
\begin{equation}
\frac{d\vec{p}}{dt} =-\eta_D(p) \vec{p} + \vec{\xi}.
\end{equation}
for the coefficients evaluated by collisional only models (\PHSD, \CpQCD, \CQPM, \Nantes), while utilizing the improved Langevin equation~\cite{Cao:2013ita}:
\begin{equation}
\frac{d\vec{p}}{dt} =-\eta_D(p) \vec{p} + \vec{\xi} + \vec{f}_{\rm gluon}.
\end{equation}
for the coefficients evaluated by collisional + radiative models (\LBT, \Duke).

Here $\eta_D \vec{p}$ is the drag force and $\vec{\xi}$ are the thermal random kicks that heavy quarks consistently receive from the medium, which satisfy $\left<\xi_i(t)\xi_j(t') \right> = \left(\kappa_L \hat{p}_i \hat{p}_j + \kappa_T (\delta_{ij} - \hat{p}_i \hat{p}_j)\right) \delta(t-t')$. In the scenario where the radiative energy loss is considered, we introduce an additional recoil force $\vec{f}_{\rm gluon}$ resulting from heavy quark emitting gluons and define it as $\vec{f}_{\rm gluon} = - d\vec{p}_{\rm gluon}/dt$. The gluon radiation probability is adopted from the higher-twist approach. More details can be found in ~\cite{Cao:2015hia,Cao:2013ita}.

The advantages of our Langevin implementation is that the interaction between heavy quarks and the medium is solely dependent on the drag and transport coefficients $\eta_D, \kappa_L, \kappa_T$, regardless of the medium degrees of freedom or the microscopic mechanism of the interaction\cite{vanHees:2004gq}. 
Therefore it is suitable to serve as a framework for the comparison of the various forms of coefficients from different models -- either calculated directly from theory, or parametrized and later estimated from experimental data. 

\subsection{\label{sec:HQ_coef} Transport coefficients}
\begin{figure*}
	\includegraphics[width=1\textwidth]{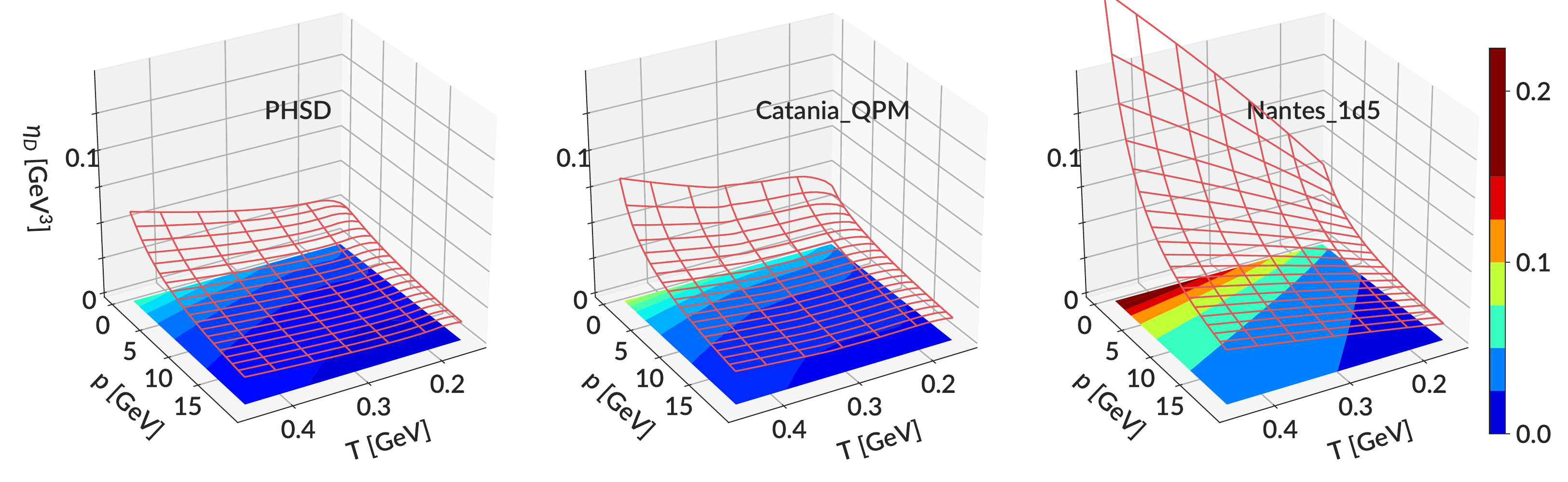}
	\includegraphics[width=1\textwidth]{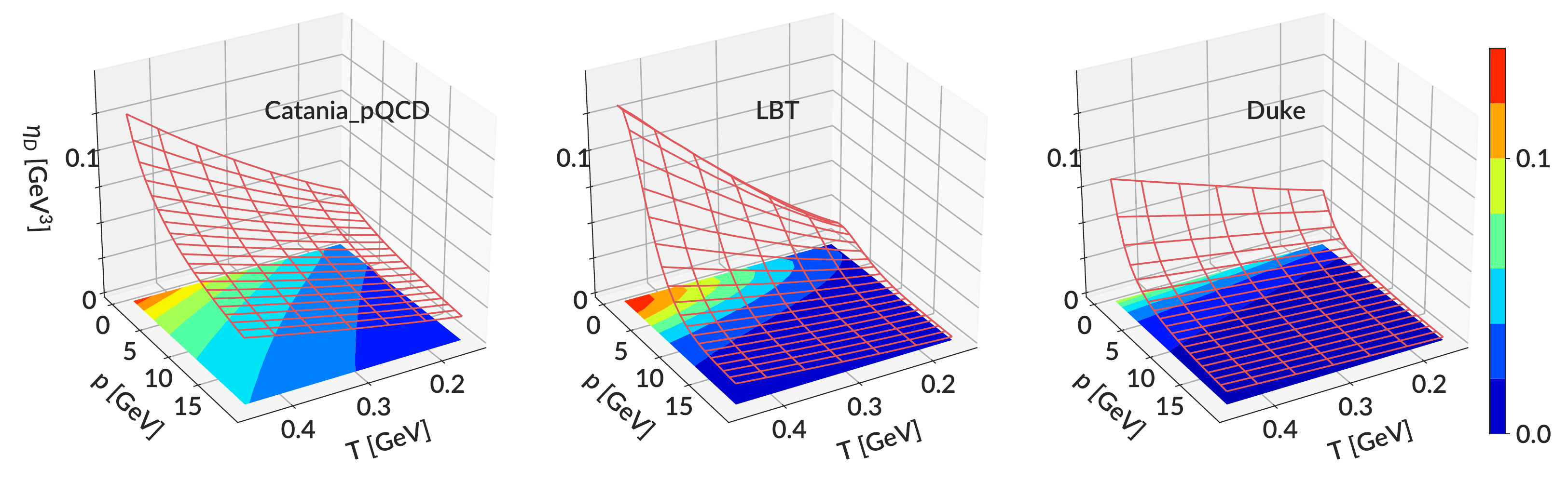}
	\caption{(Color online) Leading order charm quark transport coefficients (drag coefficients $\eta_D$) estimated by each group to describe the $D$-meson $R_{\rm AA}$ and $v_2$ at AuAu and/or PbPb collisions at RHIC and the LHC.}
\label{fig:drag}
\end{figure*}

\begin{figure*}
	\includegraphics[width=1\textwidth]{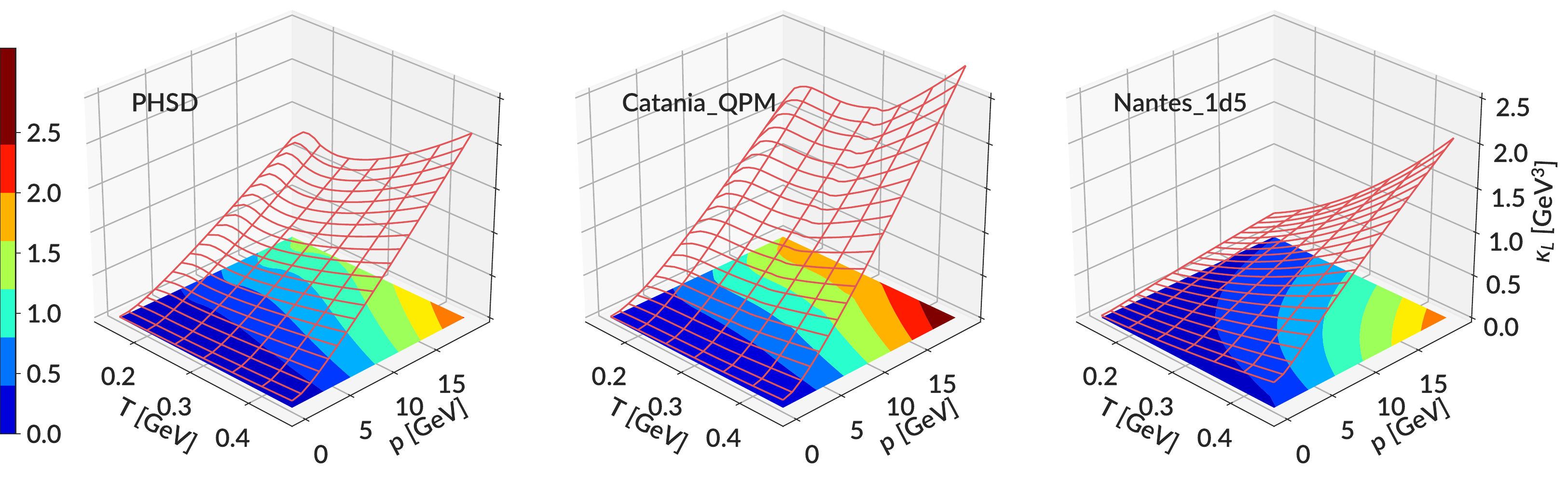}
	\includegraphics[width=1\textwidth]{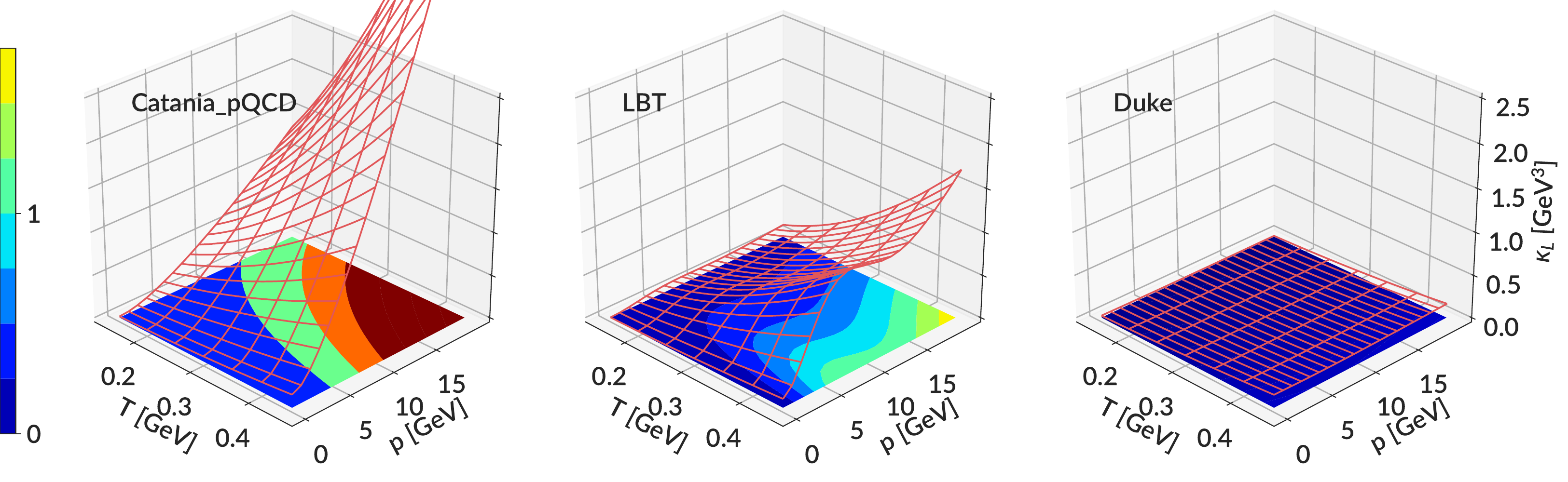}
	\caption{(Color online) Same as Fig.~\ref{fig:drag} but for longitudinal momentum transport coefficient $\kappa_L$.}
	\label{fig:kL}
\end{figure*}

\begin{figure*}
	\includegraphics[width=1\textwidth]{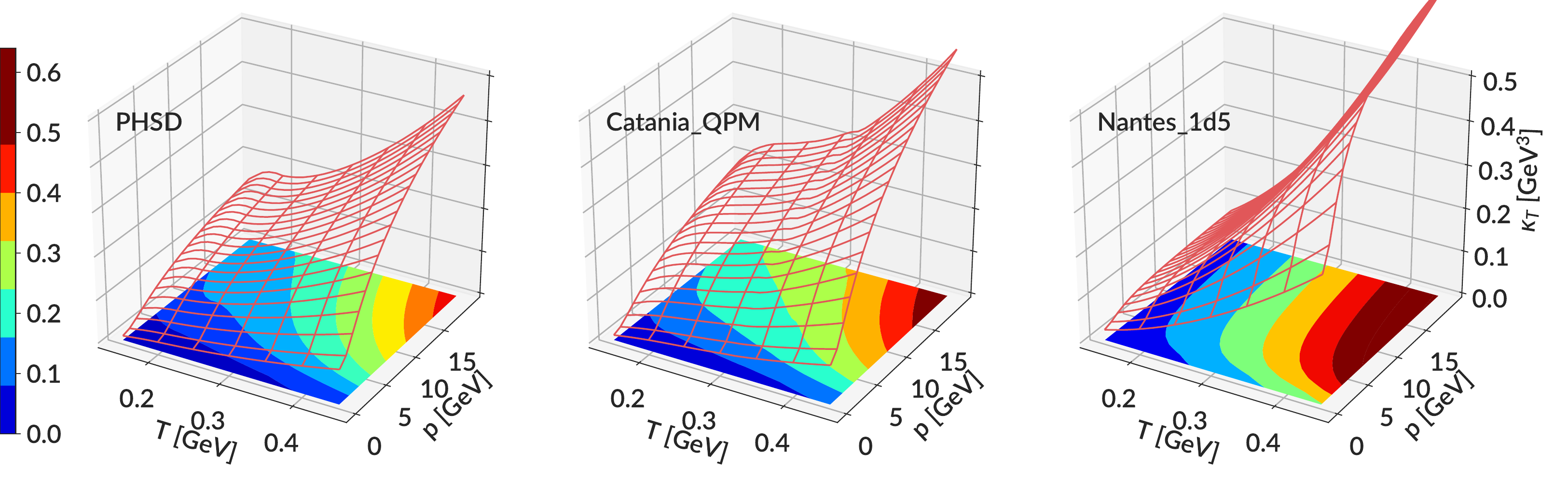}
	\includegraphics[width=1\textwidth]{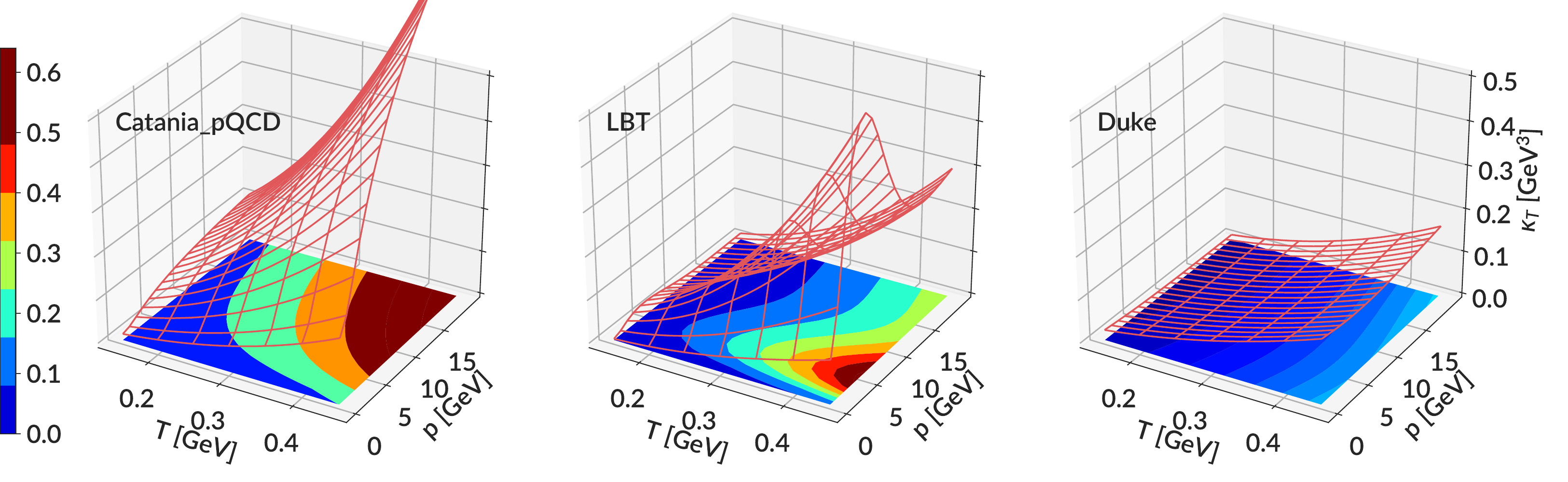}
	\caption{(Color online) Same as Fig.~\ref{fig:drag},~\ref{fig:kL} but for transverse momentum transport coefficient $\kappa_T$.}
	\label{fig:kT}
\end{figure*}
\begin{figure*}
	\includegraphics[width=1\textwidth]{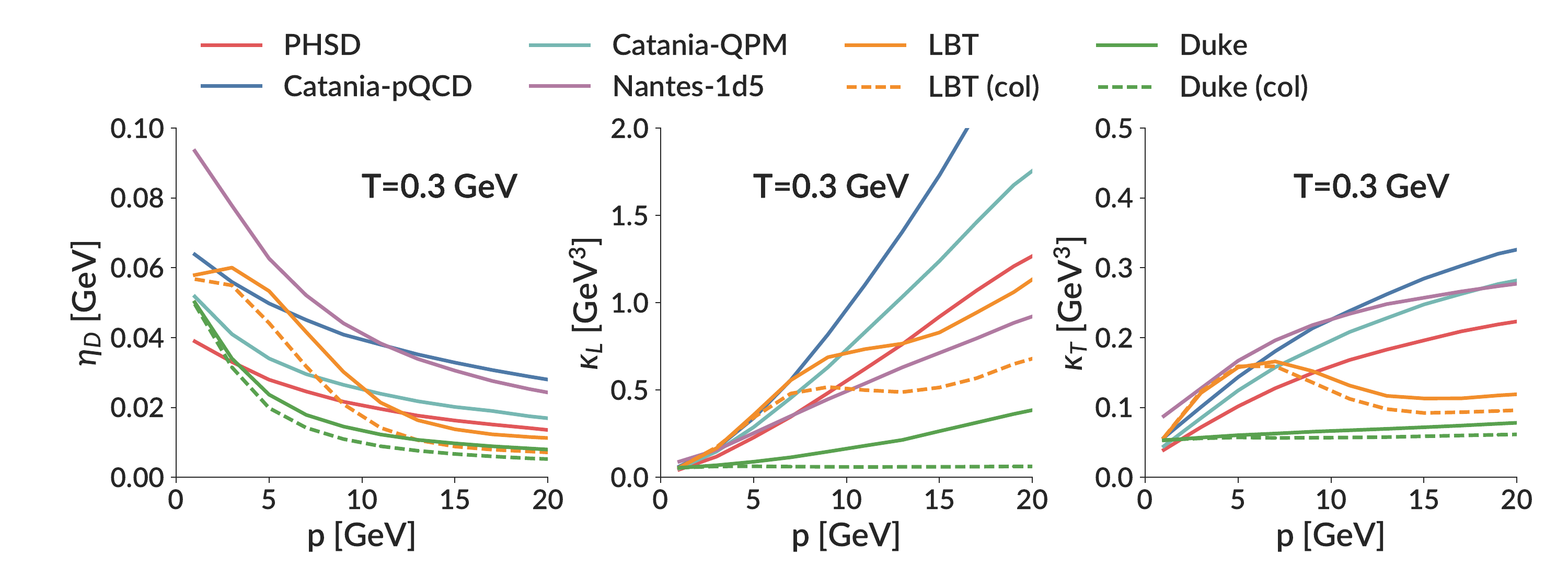}
	
	\includegraphics[width=1\textwidth]{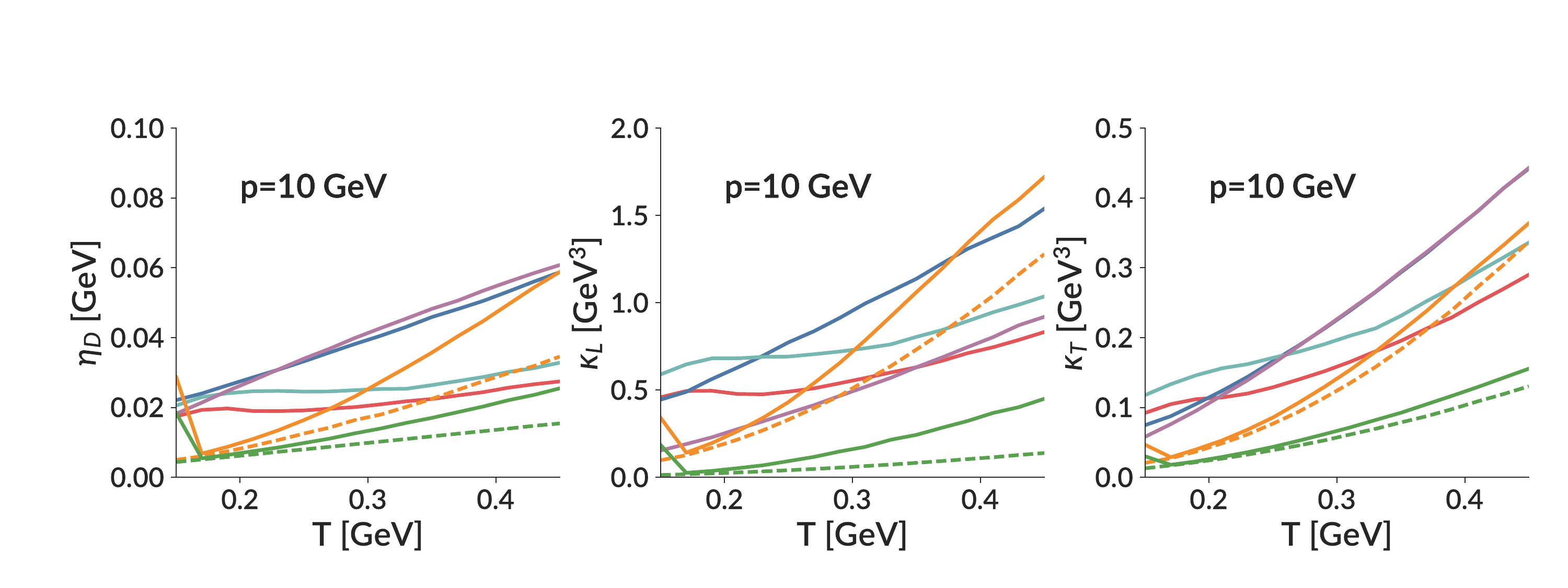}
	\caption{(Color online) Overall heavy quark transport coefficients with radiative process considered. The values are estimated by propagating charm quarks inside a static medium for 1 fm/c. The solid lines represent the total transport coefficients, while the dashed lines represent the contribution from collisional only process. The difference between those two is the contribution from higher order radiative process.}
	\label{fig:drag_dynamic}
\end{figure*}
One of the ambitious goals for heavy flavor studies in heavy-ion collisions is to get access to the properties of the QGP medium, especially to calculate or estimate the interaction between heavy quarks and the medium by encoding the interaction into a few transport coefficients.
The drag and momentum transport coefficients are defined as~\cite{Moore:2004tg}: 
\begin{equation}
	\begin{cases}
	& \frac{d}{dt}\left<p\right>  \equiv - \eta_D \left<p\right> , \\
	& \frac{1}{2} \frac{d}{dt} \left<(\Delta p_T)^2\right> \equiv \kappa_T,  \\
	& \frac{d}{dt} \left<(\Delta p_z)^2\right> \equiv \kappa_L.
	\end{cases} 
	\label{eqn:coefficients}
\end{equation}
which are the average momentum loss, the transverse and longitudinal fluctuations, respectively.

Ideally one would derive the transport coefficients through a first principle calculation and confront them directly with experimental data. 
However, most of the microscopic transport models that are applied to simulate the heavy quark evolution have approached the estimation problem more or less in a data-driven way. 
Some ad hoc parameters usually need to be introduced when implementing the heavy quark in-medium evolution, ( e.g. corrections for higher order processes or some unknown non-perturbative effects), and these parameters are later calibrated with experimental measurements. 
Clearly, part of the difference observed in the extracted transport coefficients stems from the different intrinsic interaction mechanisms that are considered when the model is implemented. 
Yet, part of the discrepancy in the transport coefficients also comes from the different choices of other components, such as initial conditions, the hadronization process, the medium evolution and so on.
All of these can have non-negligible effects on the final output, and thus in turn affect the estimation of the transport coefficients in the QGP phase.

In Fig.~\ref{fig:drag},~\ref{fig:kL},~\ref{fig:kT} we compare the charm quark drag and momentum transport coefficients $\eta_D, \kappa_L, \kappa_T$ as a function of temperature and momentum for several models listed in Sec.~\ref{sec:introduction}. 
All the transport coefficients are evaluated such that each model is able to describe the $D$-meson $R_{\rm AA}$ and $v_2$ for AuAu and/or PbPb collisions at RHIC and the LHC. 
The drag and momentum transport coefficients are separated into two groups, where the \PHSD, \CQPM, \CpQCD and \Nantes(collisional) models employ only the collisional energy loss, and the \Duke,\LBT models employ both collisional and radiative energy loss for heavy quarks.
For the drag coefficient $\eta_D$, all the models show a monotonously rising temperature dependence and a decrease for increasing momentum. The \Nantes coefficients have the largest gradient in the high temperature and low momentum region, which is due to a momentum dependent running coupling constant.

Both the transport coefficients $\kappa_L, \kappa_T$ show a strong positive momentum dependence and a mild temperature dependence, except for the \Duke coefficients, which feature an isotropy assumption unlike the others, and have the smallest absolute value. The \PHSD coefficients are consistently smaller but still compatible with the \CQPM coefficients, while some interplay appears in the low momentum region when one compares between \Nantes and \CpQCD coefficients. The non-trivial peak for \LBT coefficients in the low momentum region is due to the non-constant $K$-factor, which is included in the model in order to provide a satisfactory description of experimental data, and its parametrization reads as $K=1 + K_p \exp(-p^2/2\sigma_p^2)$.

The drag and transport coefficients shown in Figs.~\ref{fig:drag},~\ref{fig:kL},~\ref{fig:kT} carry only contributions from elastic processes. These are the most often used transport coefficients for characterizing the interaction between heavy quarks and the medium, and are frequently compared among different models.
For models that consider only collisional energy loss process, they represent the total drag and momentum coefficients. 
However, for models incorporating both collisional and radiative energy loss -- \Duke and \LBT -- the inelastic processes contribution may be significant,  even though the gluon radiation process is of higher order in $\alpha_s$. 
In order to make fair comparison between different models, it is therefore important to account for all of these contributions, from elastic and inelastic processes.

The gluon radiation implemented in \Duke and \LBT is time-dependent (proportional to $\sin^2(\Delta t/2\tau_f)$ where $\tau_f$ is the gluon formation time), which breaks the localization of the interaction and mimics the coherence effect of medium-induced radiation.
We calculate the momentum change and broadening using Monte Carlo techniques, by propagating the heavy quarks in a static medium for 1 fm/c, and calculating the total coefficients by Eqn.~\ref{eqn:coefficients}. 
The results from the dynamical calculation are presented in Fig.~\ref{fig:drag_dynamic} at fixed temperature $T=0.3$ GeV and fixed momentum $p=10$ GeV correspondingly. 
The solid lines are the overall coefficients (containing both, elastic and inelastic contributions), while the dashed lines are the contributions from elastic processes, the difference between these two are the additional contributions from inelastic processes. 
We can already see that for the \Duke and \LBT models the gluon radiation contributes effectively at higher momenta and at temperatures which we observe at the beginning of the QGP expansion. 
The existence of radiative processes can partially explain why the transport coefficients estimated by the \Duke and \LBT models are comparatively smaller than those in models containing solely elastic interactions when one only includes the elastic components in the analysis.

\section{\label{sec:in-medium} Heavy quark in-medium evolution}

In this section, we implement charm quark propagation inside a QGP medium using  Langevin dynamics coupled to a realistic description of the QGP medium in AuAu collisions at $\sqrt{s}=200 A$ GeV.
We test the impact of several model components, and compare the charm quark energy loss at the end of the QGP phase ($T_c=0.155$ GeV~\cite{Borsanyi:2010bp,Bhattacharya:2014ara}). 
The two variables evaluated are the nuclear modification factor $R_{\rm AA}$, here defined as the ratio between the final state spectra and initial state spectra $R_{\rm AA} = \frac{dN_{\rm final}}{dp_Tdy}/ \frac{dN_{\rm initial}}{dp_Tdy}$, and the elliptic flow $v_2 = \left< \frac{p_y^2 - p_x^2}{p_y^2 + p_x^2} \right>$. We do not intend to compare the different hadronization mechanisms , which are among the least understood processes yet have been investigated in \cite{Rapp:2018qla}.

This section is structured as follows: first we will compare the results generated from different initial conditions using the same QGP medium evolution model (a hydrodynamical description), and the heavy quarks interact with the medium using common transport coefficients.
Then we will compare the results using different medium evolution models, in which the media expand from the same initial conditions.
Later we compare in detail how heavy quarks respond to different drag and momentum transport coefficients for a given standard initial condition and a common medium evolution.
Different schemes for the Einstein relationship implementation as well as the temperature and momentum dependence of the drag and transport coefficients are inspected at the end of this section.

\subsection{\label{subsec:IC}Initial conditions}
\begin{figure*}
	\includegraphics[width=\linewidth]{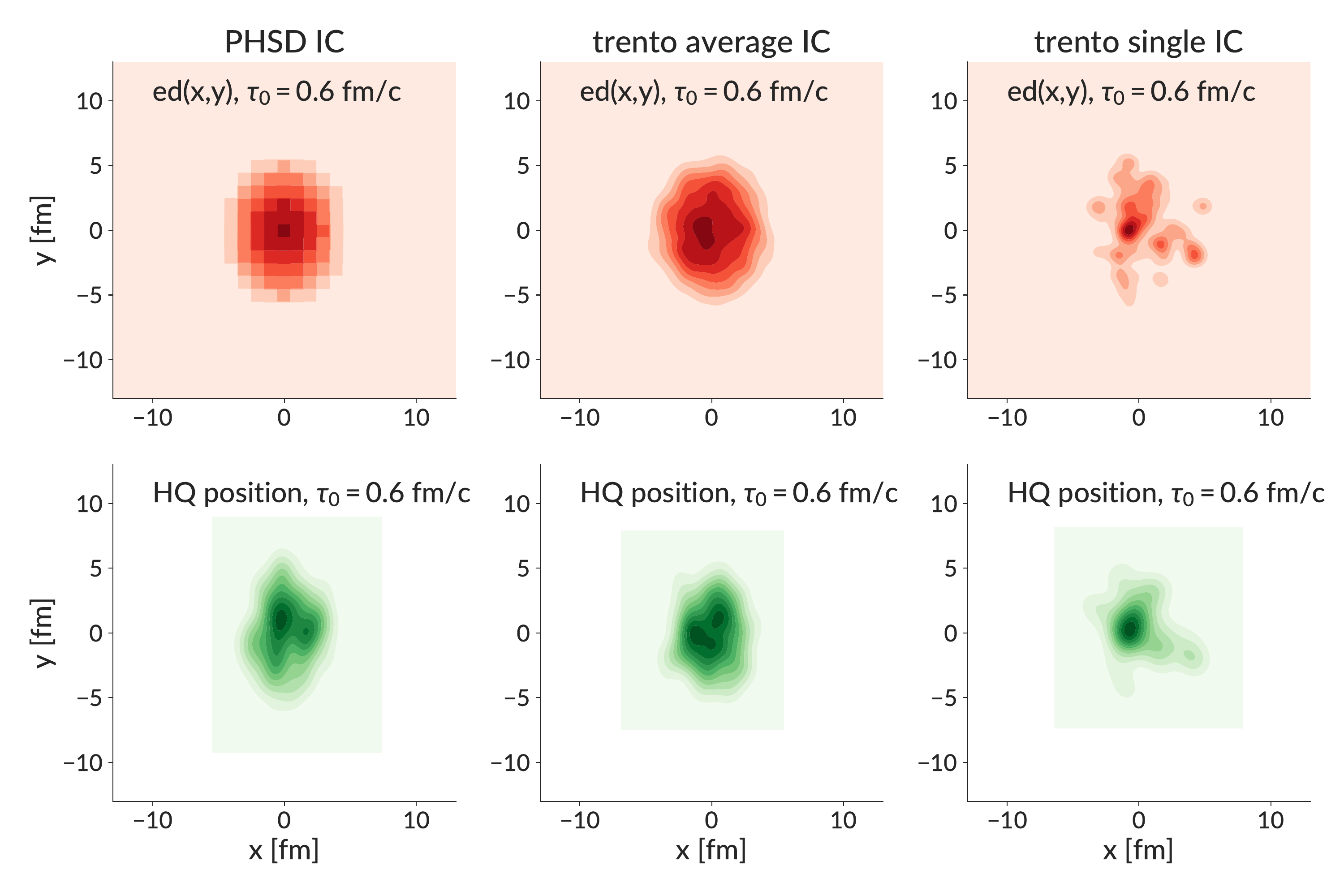}
	\caption{(Color online) Initial conditions at $\tau_0=0.6$ fm/c for AuAu collisions at 200 GeV with impact parameter $b=6$ fm, generated from: (Left): PHSD initial conditions; (Middle): trento average initial condition, which is averaged by 50 trento events who share the similar spatial eccentricity as in PHSD initial condition; (Right): one example of single trento event.}	
	\label{fig:IC_density}
\end{figure*}

\begin{figure*}
	\begin{minipage}{.37\textwidth}
		\includegraphics[width=\linewidth]{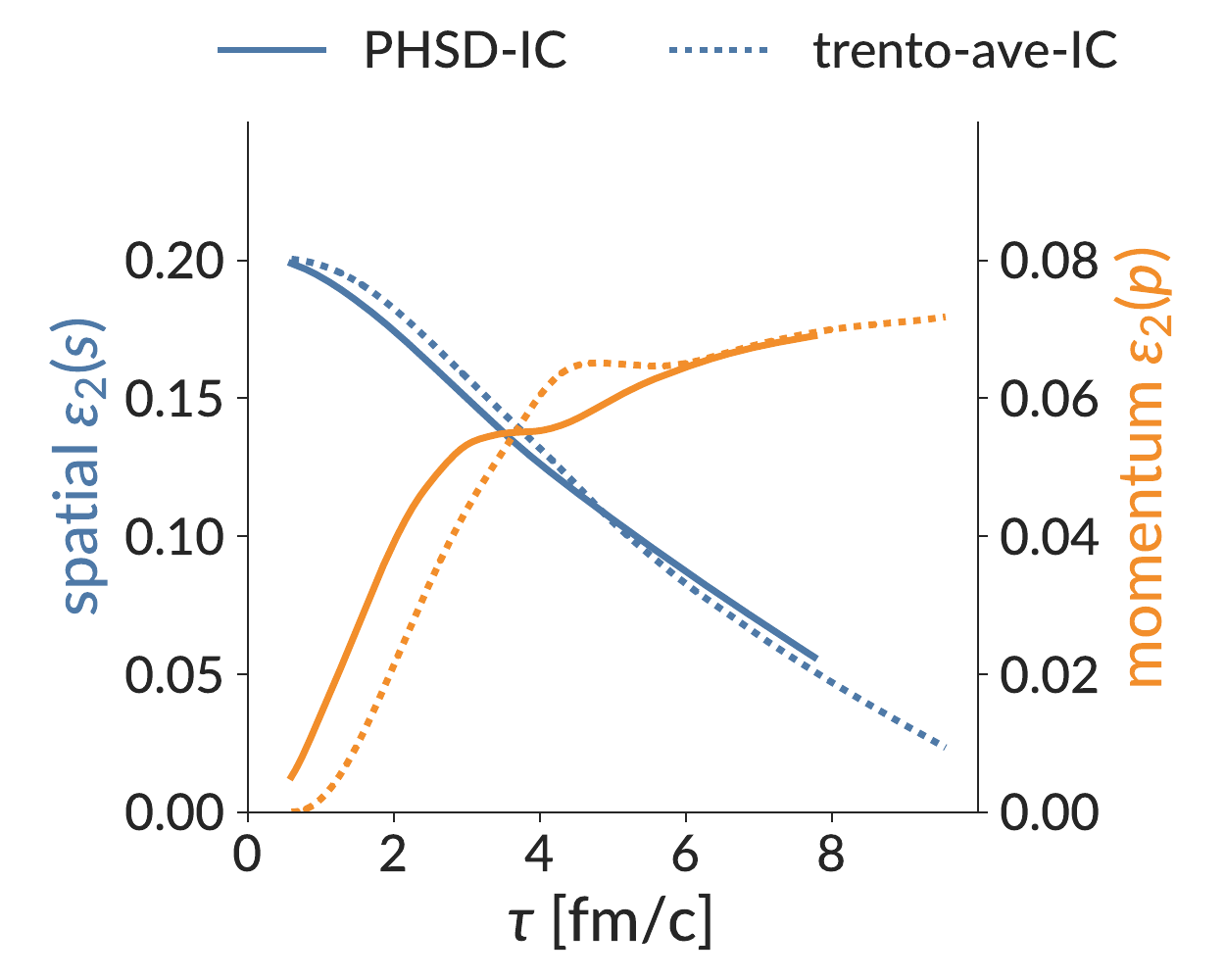}
		\caption{(Color online) Time evolution of the spatial and momentum anisotropy of the QGP medium that is simulated by a (2+1)D hydrodynamical model -- VISHNU\cite{Song:2007fn,Qiu:2011hf}. The medium starts from two different initial conditions: PHSD initial condition, and averaged trento initial condition.}
		\label{fig:eccen_soft}
	\end{minipage} \quad
	\begin{minipage}{.60\textwidth}
		\includegraphics[width=\linewidth]{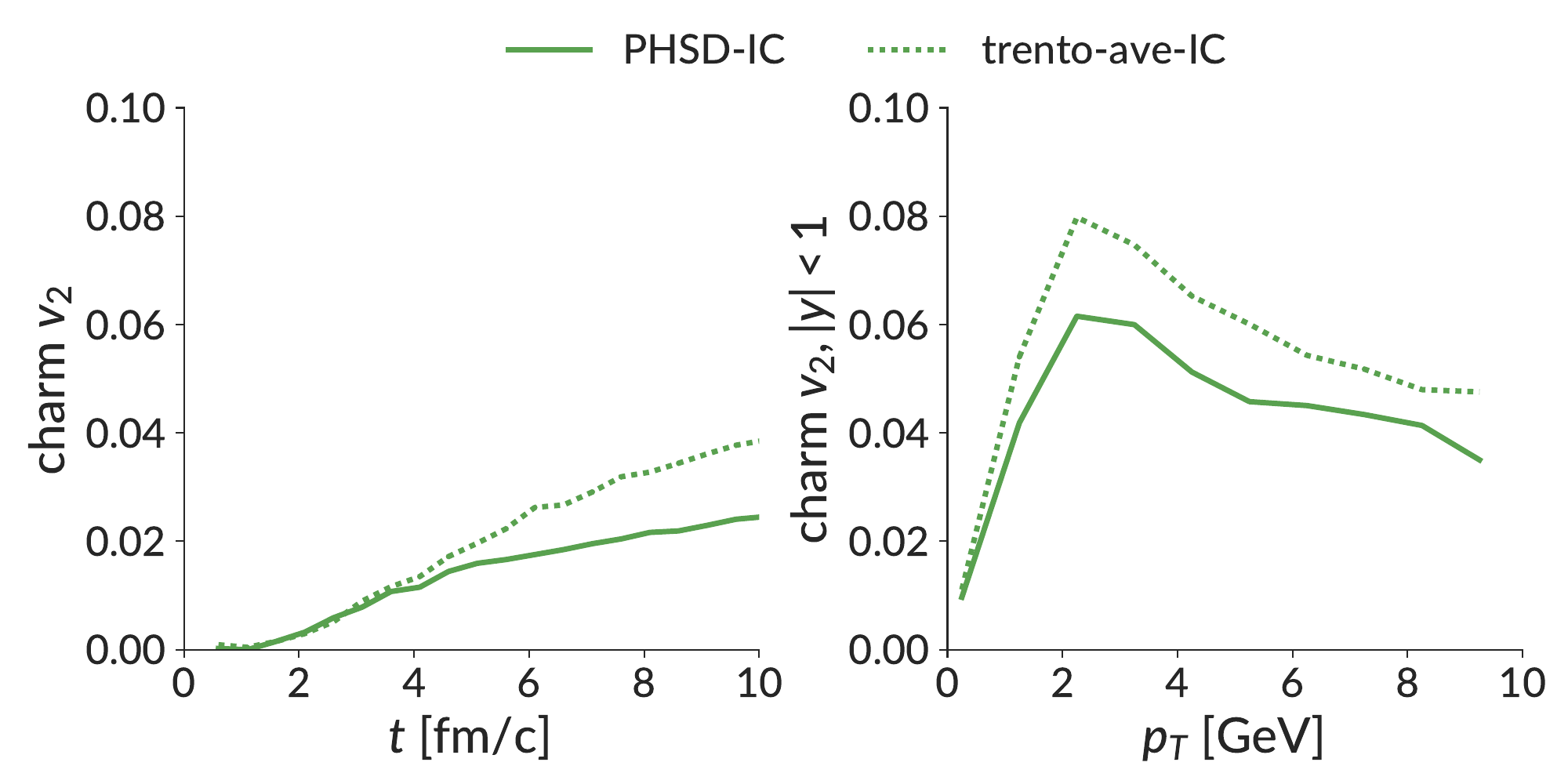}
		\caption{(Color online) (\textbf{Left}): Development of charm quark elliptic flow inside a (2+1)D hydrodynamical medium during the QGP phase. (\textbf{Right}): charm quark $v_2(p_T)$ evaluated at the end of the QGP phase. The charm quarks interact with the medium following an improved Langevin dynamics, with \Duke coefficients applied.}
		\label{fig:HQ_v2_IC}
	\end{minipage}
\end{figure*}

Charm quarks are created initially by hard processes which can be calculated in perturbative QCD. 
In this study we employ FONLL~\cite{Cacciari:1998it,Cacciari:2012ny} complemented by the nuclear shadowing effect (cold nuclear matter effect) in the EPS09 parametrization~\cite{Eskola:2009uj} to calculate the initial spectra of the c-quarks.

In position space the initial geometry of the collisions still remains one of the largest uncertainties in modeling the QGP evolution in heavy-ion collisions~\cite{Schenke:2013dpa}. Therefore the correlation between the initial energy/entropy density of the QGP and the initial position distribution of heavy quarks is still a matter of active research.

In this study, we compare two different initial conditions:
\begin{itemize}
	\item \PHSD:  the procedure developed in Ref. \cite{Xu:2017pna} is applied in order to solve the Landau matching condition $T^{\mu\nu}u_\nu = e u^\mu$ by diagonalization of the energy-momentum tensor extracted from the PHSD simulations. We construct a grid in Milne coordinates with a cell size $\Delta \tau = 0.2$ fm/c, $\Delta x = \Delta y = 1\ \text{fm}$ and $\Delta \eta = 0.1$).
 The starting time $t = 0$ considered here corresponds to the first nucleon-nucleon collision. In the PHSD model, the particle coordinates are converted from ($t,x,y,z$) to Milne coordinates using the following relations ($\tau = \sqrt{t^2-z^2},x,y,\eta = 1/2 \ln((t+z)/(t-z))$). Each of these particles then contributes to the energy-momentum tensor with a Gaussian weight where the widths are taken to be $(\Delta x)^2 = (\Delta y)^2$ in the transverse direction and $(\Delta \eta)^2$ in the longitudinal direction. This smearing procedure allows us to obtain a smooth profile with only one PHSD simulation containing 30 parallel events. To avoid any over-counting, each particle is restricted to only contribute once in a given bin in proper time $\tau$. In addition, particles with non-real proper times and space-time rapidities are simply discarded. Using this method, the local energy density $e$, the pressure components and the cell flow velocity $\vec{\beta}$ are extracted for each space-time cell of the Milne grid. 
	
	\item \trento~\cite{Moreland:2014oya}: a parametric initial condition that does not assume a specific physical mechanism, but deposits energy/entropy according to a parametric function that maps the projectile thickness $T_A$ into initial distribution $dS/dy$ at mid-rapidity. The mapping function is calibrated to experimental data of light hadron observables by Bayesian analysis and the functional form $dS/dy \propto \sqrt{T_A T_B}$ is used in this comparison.
\end{itemize}
Although initial event-by-event fluctuations are generally regarded as an important feature in modeling the collision and have been shown to have a considerable impact on flow observables, here, we will for the sake of simplicity, consider averaged \trento initial conditions which are obtained using 50 single \trento events. These averaged initial conditions are widely utilized in hydrodynamical models in the literature and are computationally significantly less expensive.

Figure~\ref{fig:IC_density} shows a PHSD initial condition, a \trento initial condition and an averaged \trento initial condition for AuAu collisions at$\sqrt{s}= 200$ AGeV with an impact parameter $b=6$ fm at the hydro starting time $\tau_0=0.6$ fm/c (as well as the starting time of heavy quarks interacting with the medium).
The top figures are the initial energy density for the soft medium, while the bottom figures are the histograms of initial heavy quark positions for the corresponding same events.
The (averaged) \trento initial condition is constructed by averaging over 50 independent \trento initial conditions. All the \trento  initial conditions are selected to have a similar spatial eccentricity $\epsilon_2(s)$ as the PHSD initial condition. 
Those initial energy densities are the input for the (2+1)D hydrodynamical model -- {\tt VISHNU} -- to simulate the evolution of the QGP medium, starting from $\tau_0=0.6 fm/c$. Figure~\ref{fig:eccen_soft} (left), shows the time evolution of the spatial and the momentum eccentricity of the medium, displaying the well-known behavior of decreasing $\epsilon_2(s)$ and increasing $\epsilon_2(p)$ as the system expands.
The momentum eccentricity can be interpreted as the response of the system to the initial spatial eccentricity. 
The hydrodynamic medium evolution with the PHSD initial condition shows a more rapidly increasing momentum anisotropy at earlier times of the evolution (due to the initial flow $\vec{\beta}$ introduced in the system) and slowing down after the first few fm/c. The final momentum anisotropy, however, is comparable to the one with an averaged \trento initial condition.

Using these initial conditions, we then propagate charm quarks in the QGP medium till the end of the QGP phase, using \Duke transport coefficients (this choice is arbitrary - we just need to fix one set of coefficients for the comparison). 
Figure~\ref{fig:HQ_v2_IC} (left), shows the time evolution of the elliptic flow of charm quarks. A significant fraction of this elliptic flow is generated at later times during the evolution, when the medium itself has a larger momentum anisotropy.
We observe a persistent difference between the charm quark $v_2$ generated by these two different initial conditions. This implies that charm quarks can actually not only retain information about the initial condition, but also keep a record of the QGP medium expanding history, particularly, the later stages of the evolution. 
At the end of the QGP phase, charm quarks starting from an average \trento initial condition have picked up a larger $v_2$ than the ones from the PHSD initial conditions, as shown on the right of Fig.~\ref{fig:HQ_v2_IC}, plotted as the charm quark $p_T$ differential flow at the end of the QGP phase.

\subsection{\label{subsec:medium}QGP medium evolution}
\begin{figure}
	\includegraphics[width=0.5\textwidth]{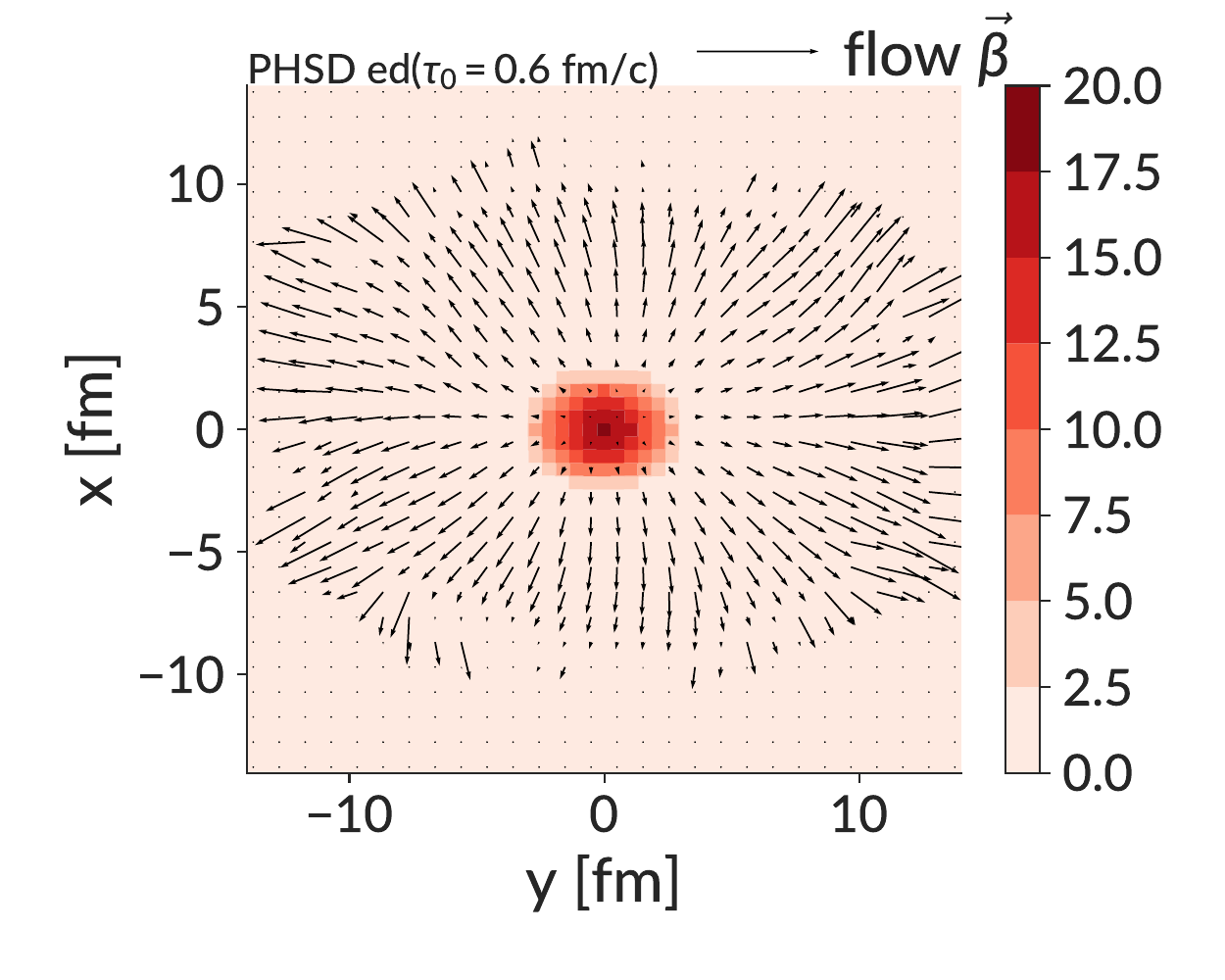}
	\caption{(Color online) Hydro initial condition ($\tau_0=0.6$ fm) generated from PHSD, used as input for hydro evolution.}
	\label{fig:PHSD_IC}
\end{figure}

\begin{figure*}
	\includegraphics[width=1.\textwidth]{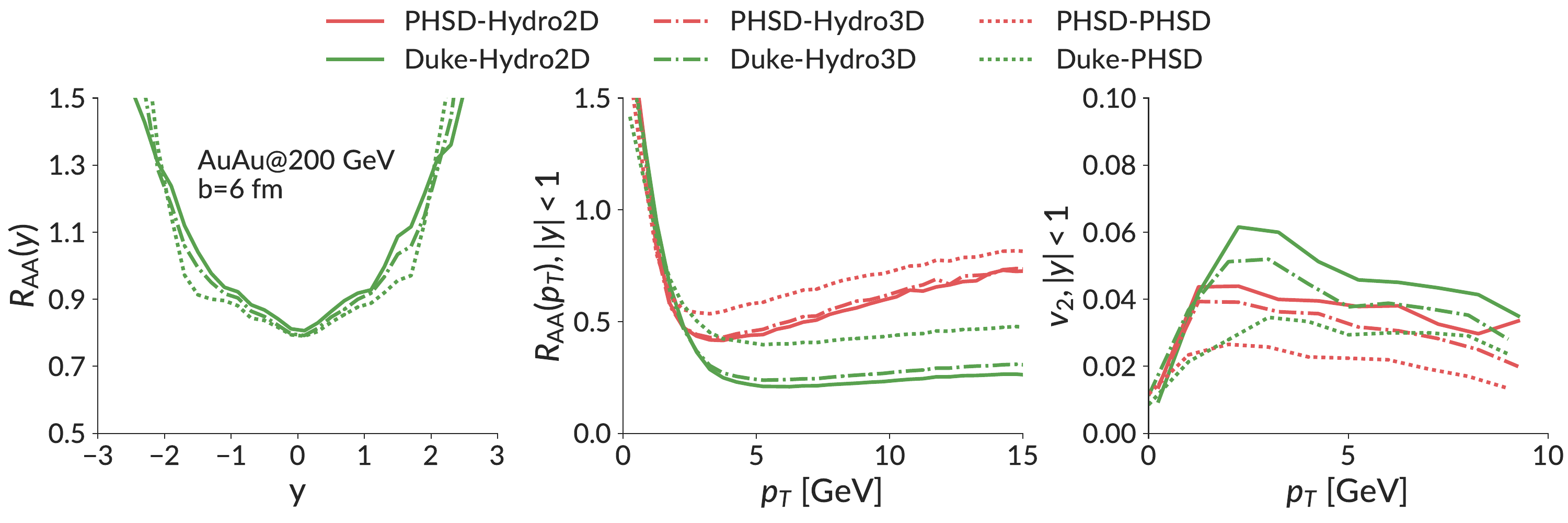}
	\caption{(Color online) Charm quark $R_{\rm AA}$ as a function of $y$ (\textbf{Left}), and $p_T$ (\textbf{Middle}), elliptic flow $v_2$ as a function of $p_T$ (\textbf{Right}) at the end of the QGP phase. The charm quarks are following a Langevin dynamics with two sets of transport coefficients applied: \PHSD (red) and \Duke (green).}
	\label{fig:3medium_b6fm}
\end{figure*}
The interaction between heavy quarks and the medium is dependent on the local temperature and the flow velocity of the medium.
In microscopic transport models such as Boltzmann dynamics, the interactions will also depend on the medium degrees of freedom. 

Various approaches have been developed to describe the evolution of the QGP medium in heavy-ion collisions. 
Hydrodynamical models are very successful in describing hadron multiplicities, and flow coefficients up to $p_T\simeq 3$ GeV. 
However, they do rely on the assumption of local thermal equilibrium during the evolution. 
An alternative approach to the hydrodynamical description is microscopic transport, which employs microscopic kinetic theory and evolves the system of partons using a transport equation such as the Boltzmann equation~\cite{Uphoff:2012gb,Song:2015sfa,Scardina:2017ipo}. 
These types of models do not rely on any equilibrium assumptions but require assumptions on the medium degrees of freedom. 
The differences between those two classes of models result in significant deviation of the medium properties, especially the viscosity corrections. 
An early comparison between a hydrodynamical model and an expanding fireball model already revealed some significant differences regarding the charm quark $v_2$ at the end of QGP phase due to the different development of radial and elliptic flow in those two models~\cite{Gossiaux:2011ea,Alberico:2011zy}. 
Thus one may also expect differences if one compares medium evolutions based on hydrodynamics vs. kinetic theory.

Here we briefly summarize the default medium evolution models that are utilized in the heavy quark transport models mentioned in Sec.~\ref{sec:introduction}.

\begin{itemize}
\item PHSD: off-shell transport approach with a hadronic and a partonic phase, the simulation of the medium is based on the off-shell Kadanoff-Baym equations (in first order gradient expansion), the medium consists of quasi-particles, whose masses and widths are determined by fitting the lattice QCD EoS. (\PHSD)
\item (2+1)D viscous hydrodynamical model VISHNU: implementation of boost-invariant viscous hydrodynamics which has been updated to handle event-by-event fluctuated initial conditions and incorporates shear and bulk viscosity corrections with temperature dependence, calculating the second-order Israel-Stewart equations in the 14-momentum approximation (\Duke, \LBT).   
\item Boltzmann transport model~\cite{Plumari:2012ep}: full Boltzmann simulation with QGP medium composed either of pQCD massless or massive particles. The local cross section for the interaction between the bulk constituents is tuned to a fixed value of $\eta/s(T)$. This is realized through the Chapmann-Enskog approximation and allows to gauge the Boltzmann collision integral to the wanted $\eta/s(T)$ and simulate the fluid evolution in analogy to hydrodynamic approach.(\CpQCD , \CQPM)
\item EPOS: event generator with fluctuating initial conditions and (3+1)D viscous hydrodynamics {\tt vHLLE} using a lattice QCD EoS. (\Nantes)
\end{itemize}

In this section we compare the charm quark propagation through three different QGP medium evolutions: the PHSD medium, the (2+1)D hydrodynamical VISHNU medium, and the (3+1)D hydrodynamical vHLLE medium~\cite{Karpenko:2013wva}. 
We prepare the medium evolutions following the same methodology as discussed in~\cite{Xu:2017pna}, starting from the same initial conditions (initial energy density $e$, flow velocity $\vec{\beta}$) that have been generated by the PHSD model. The initial energy density and transverse flow $\vec{\beta}$ generated from PHSD is plotted in Fig.~\ref{fig:PHSD_IC}, at hydro starting time $\tau_0=0.6$ fm/c.
A detailed comparison regarding the PHSD medium and the hydrodynamical medium can be found in~\cite{Xu:2017pna}.

The charm quarks then propagate through the three media using our standard Langevin dynamics, where two sets of transport coefficients are chosen as examples: the collisional-only \PHSD coefficients, and the collisional + radiative \Duke coefficients. 
Charm quark $R_{\rm AA}(y), R_{\rm AA}(p_T)$ and $v_2$ are evaluated at the end of the QGP phase, and are shown in Fig.~\ref{fig:3medium_b6fm}.

As shown in Fig.~\ref{fig:3medium_b6fm} the evolution of charm quarks inside hydrodynamical (2+1)D {\tt VISHNU} and (3+1)D {\tt vHLLE} media are quite similar to each other. 
For the $R_{\rm AA}$ with respect to rapidity $y$, whose value is dominated by low $p_T$ charm quarks, discrepancies among the three media appear at large rapidities. 
Among those, the low $p_T$ charm quarks are most suppressed in a PHSD medium around $|y|\simeq 2$.  

High $p_T$ charm quarks propagating inside a hydrodynamical medium (solid and solid dots lines) show a larger suppression than in the PHSD medium (dots lines)
and develop a larger elliptic flow $v_2$. While  $R_{\rm AA}(y)$ and $R_{\rm AA}(p_T)$ are almost identical for (2+1)D and (3+1)D hydrodynamical calculations, the values of $v_2$ differ by about 15\%.
This  is understandable as the medium anisotropy is weaker in a (3+1)D simulation but also reveals the limitation of the predictive power of (2D+1) hydrodynamical calculations .
The difference between charm quarks propagating in a hydrodynamical medium and a PHSD medium, however, is more significant. A factor of 2 difference in the momentum differential flow $v_2$ is observed in the high momentum region.

A previous study~\cite{Xu:2017pna} has shown that although the shear (bulk) viscosity implemented in the hydrodynamical medium are compatible (smaller) than what is embedded in the PHSD model, the latter has a weaker response to the bulk pressure, resulting in a slightly smaller momentum eccentricity for the bulk sector at later times of the evolution in the PHSD model.
Recalling what is shown in previous section, charm quarks develop a significant part of their flow at later evolution times.
The substantial discrepancy between the charm quark evolution inside the two different media, shows that charm quarks are more susceptible to the different bulk pressures of the media, compared to the bulk matter itself.
This study shows that the heavy quark observables are sensitive to both, the heavy quark - medium interaction and the description of the QGP expansion. One of the caveats is that different combinations of the transport coefficients and the medium expansion can lead to very similar results in one observable -- for example, the charm quark $v_2$ results of the PHSD (coefficients)-Hydro3D (medium) combination and Duke (coefficients)-PHSD (medium) , whereas for other observables, like $R_{\rm AA}(p_T)$,  the results are rather different. This reveals that multiple (additional) observables are necessary to  uniquely determine the transport coefficients and the medium expansion even if all other ingredients, like the initial conditions, were known.

\subsection{\label{subsec:coefficient}Heavy quark transport coefficients}
\begin{figure*}
	\includegraphics[width=1\textwidth]{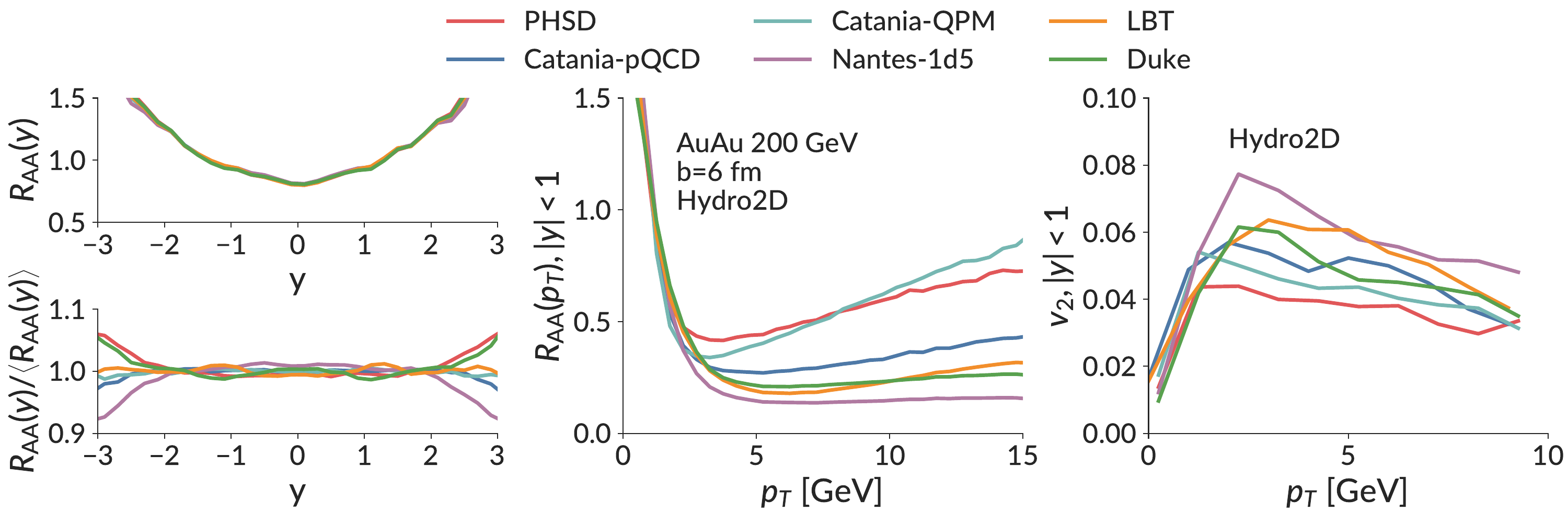}
	\includegraphics[width=1\textwidth]{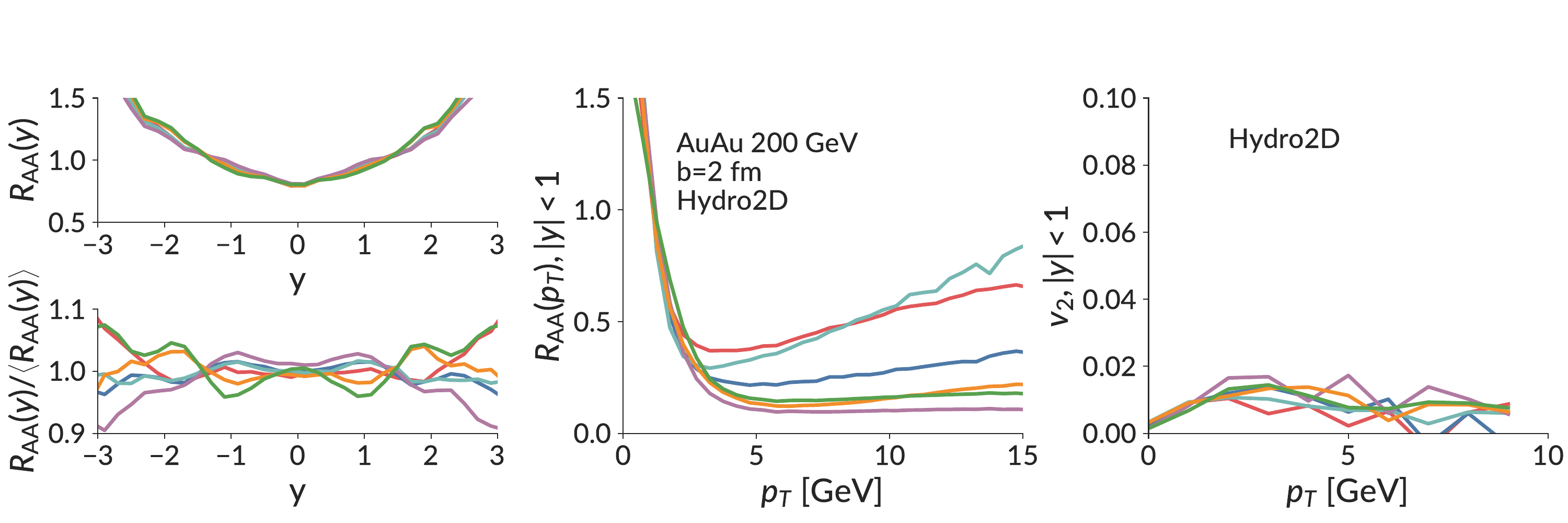}
	\caption{(Color online) Charm quark $R_{\rm AA}$ as a function of $y$ (\textbf{Left}), and $p_T$ (\textbf{Middle}), elliptic flow $v_2$ as a function of $p_T$ (\textbf{Right}) at the end of the QGP phase. The charm quarks are propagating in a hydrodynamical medium simulation for AuAu collisions at 200 GeV with $b=6$ fm/c (\textbf{Upper}), and $b=2$ fm/v (\textbf{Bottom}).}
	\label{fig:Raa_v2_6fm}
\end{figure*}
The interactions between heavy quarks and the medium are encoded into transport coefficients, which have a non-trivial temperature and momentum dependence.

In this section, we will implement different sets of charm quark transport coefficients into our standard Langevin dynamics coupled to the same  (2+1)D hydrodynamical medium, evolved from the same PHSD initial conditions. 
This setup will not only provide us with a direct comparison between the response of the charm quark observables ($R_{\rm AA}$ and $v_2$) to the transport coefficients, but also give us an insight into the difference of the interaction mechanisms employed by each model, in particular, Langevin dynamics versus microscopic transport dynamics.

The results of charm quark $R_{\rm AA}$ and $v_2$ at the end of the QGP phase are plotted in Fig.~\ref{fig:Raa_v2_6fm}. 
At intermediate and higher $p_T$ ($>5$ GeV), notable differences appear among different sets of coefficients. 
The \PHSD and \CQPM models have very similar transport coefficients, and therefore their $R_{\rm AA}$ and $v_2$ are comparable to each other. Both generate the least suppression and the smallest momentum anisotropy. 
The $R_{\rm AA}$ also levels off at higher $p_T$ due to the lack of radiative energy loss.

The \Duke and \LBT coefficients result in  moderate suppression and flow among the six, although the \Duke coefficients are the smallest of all. 
This a the consequence of including the radiative energy loss in the improved Langevin equation, which significantly strengthens the interaction between heavy quarks and the medium. 
The \Nantes coefficients result in the strongest suppression and the largest flow, even though the \Nantes($\kappa_L, \kappa_T$) are not the largest. 
In fact, when one examines the \Nantes and \CpQCD coefficients, these two are comparable with each other yet the $R_{\rm AA}$ and $v_2$ are substantially different.  
This could be a consequence of the stronger momentum dependence of the drag coefficient $\eta_D$ presented in the \Nantes coefficients, which results in a greater energy loss in a dynamical medium.

The $R_{\rm AA}$ with respect to rapidity, which is dominantly driven by the low $p_T$ charm quarks, has less differentiating power in terms of different transport coefficients. However, the rapidity dependence of heavy charm observables may still be useful for distinguishing features in the medium evolution, as demonstrated in Sec.~\ref{subsec:medium}.

At the end of this section, we should be cognizant that the observed variability of $R_{\rm AA}(p_T)$ and $v_2(p_T)$ resulting from different description of the medium expansion (as shown in Fig.~\ref{fig:3medium_b6fm}), is of the same order of the magnitude as the variability resulting from different sets of transport coefficients (Fig.~\ref{fig:Raa_v2_6fm}). Clearly one approach to improve upon this particular ambiguity is to make sure that the respective medium evolution is calibrated to well reproduce the largest possible set of observables in the light hadron sector.

\subsection{Einstein's relationship}
\begin{figure*}
	\includegraphics[width=1\textwidth]{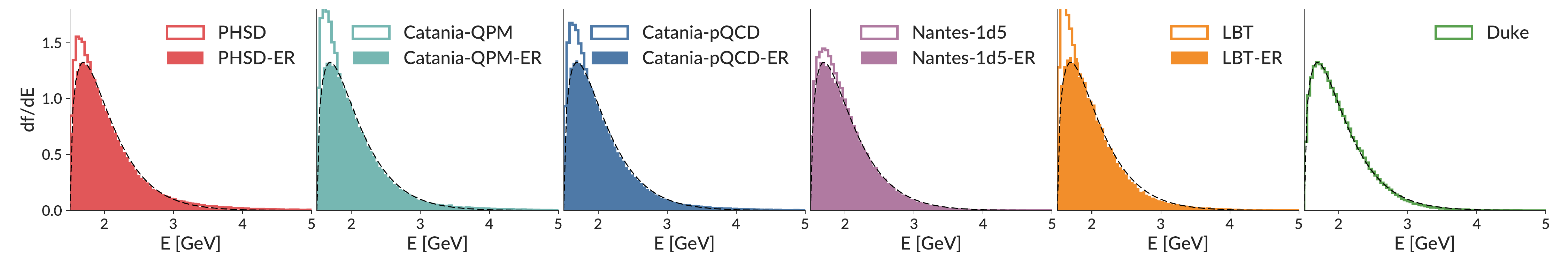}
	\caption{(Color online) Charm quarks distribution after propagating within the Langevin evolution in a static medium with constant temperature for $t=50$ fm/c. The dashed black lines are the equilibrium Boltzmann distribution.}
	\label{fig:dist_ER}
\end{figure*}

\begin{figure}
	\includegraphics[width=0.5\textwidth]{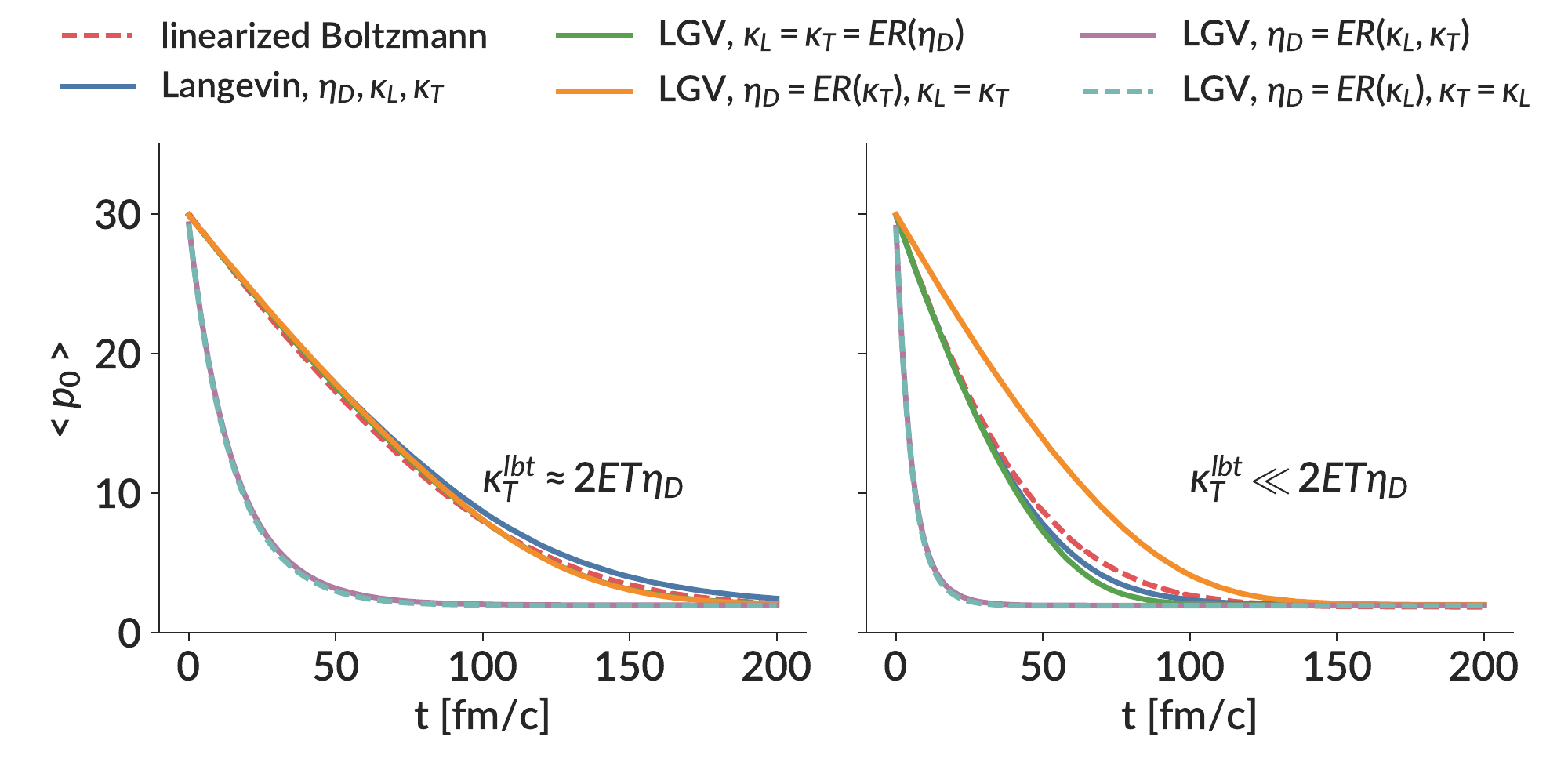}
	\caption{(Color online) Comparison between different implementation of Einstein's relationship in a static medium. Heavy quarks are initialized with same momentum as $p_z(0)=30$ GeV, and propagate in a static medium ($T=0.3$ GeV) for 200 fm/c under: linearized Boltzmann dynamics (red dash line), Langevin dynamics taking ($\eta_D, \kappa_L, \kappa_T$) calculated from linearized Boltzmann dynamics (blue line), Langevin dynamics taking ($\kappa_T$) while $\kappa_L, \eta_D$ are calculated from Einstein's relationship (orange line), Langevin dynamics taking ($\kappa_L$) while $\kappa_T, \eta_D$ are calculated from Einstein's relationship (cyan line), Langevin dynamics taking ($\kappa_T, \kappa_L$) while $\eta_D$ are calculated from Einstein's relationship (purple line), Langevin dynamics taking ($\eta_D$) while $\kappa_L, \kappa_T$ are calculated from Einstein's relationship (green line).}
	\label{fig:LBT_vs_LGV}
\end{figure}

\begin{figure*}
	\includegraphics[width=1\textwidth]{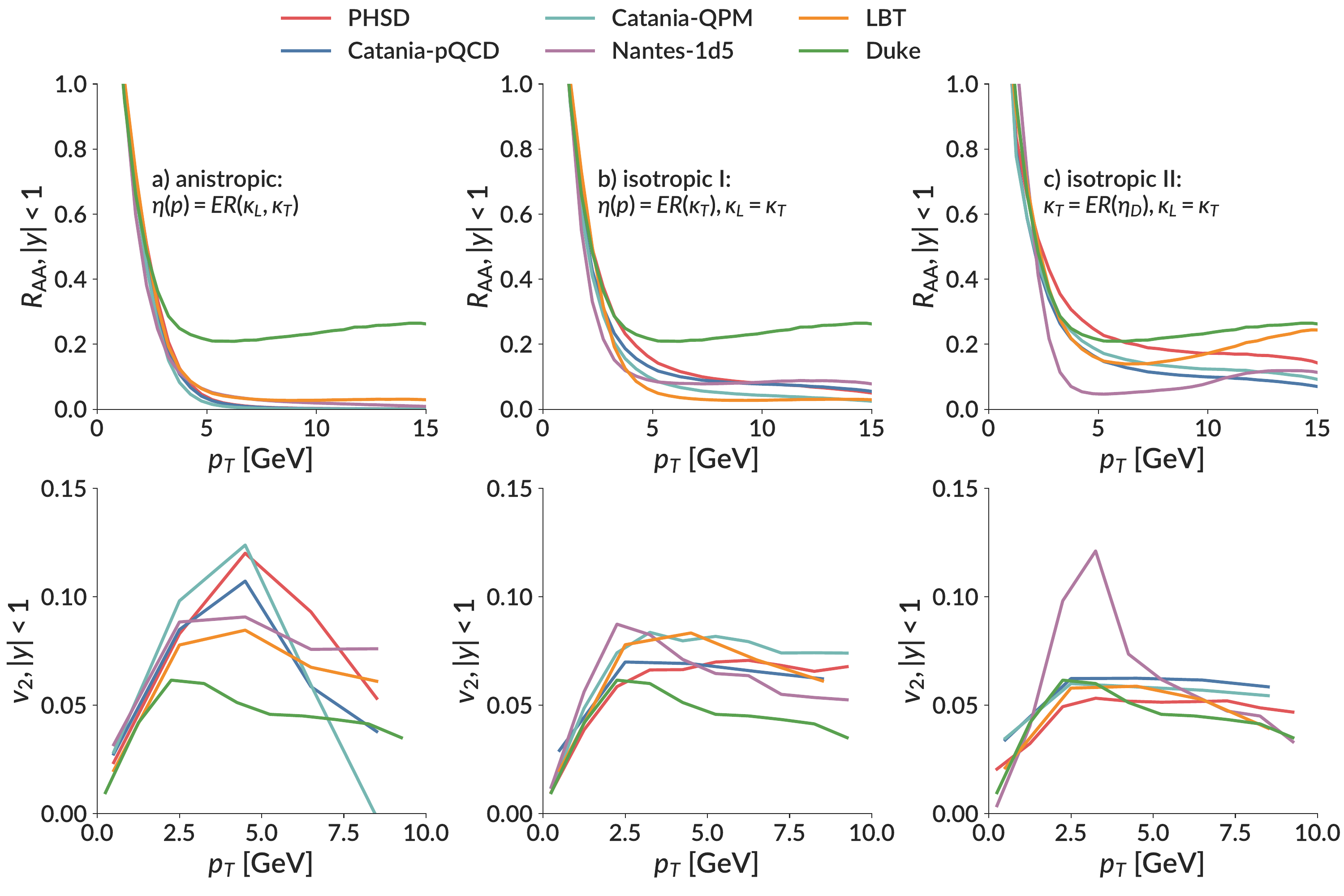}
	\caption{(Color online) Charm quark $R_{\rm AA}$ as a function of $p_T$ (\textbf{Upper}), elliptic flow $v_2$ as a function of $p_T$ (\textbf{Bottom}) at the end of the QGP phase. The charm quarks are propagating in a hydrodynamical medium simulation for AuAu collisions at 200 GeV with $b=6$ fm/c, with different scenario of Einstein's relationship applied.}
	\label{fig:Raa_v2_6fm_ER}
\end{figure*}

Most of the transport approaches presented here are based on the Boltzmann equation. The Boltzmann collision integral is solved by a Monte Carlo procedure which conserves energy and momentum for each collision. The drag and momentum transport coefficients can be calculated from the Boltzmann collision integral \cite{Berrehrah:2014tva}. 
The H-theorem guaranties that a system approaches thermal equilibrium when its time evolution is given by the Boltzmann equation.
However, when using these transport coefficients in a Langevin equation, the approach to equilibrium is only guarantied when the Einstein's relationship between the drag and the momentum transport coefficients is satisfied~\cite{vanHees:2005wb,Gossiaux:2006yu,Das:2010tj}. When the equilibrium distribution obeys Boltzmann-J\"{u}ttner statistics, the Einstein's relationship yields (in the pre-point discretization scheme of the Langevin equation): 
\begin{equation}
\eta_D = \frac{\kappa_L}{2 ET} - \frac{\kappa_L - \kappa_T}{p^2} - \frac{\partial \kappa_L}{\partial p^2}
\label{eqn:ER}
\end{equation}
 
The drag and momentum transport coefficients determined by the Boltzmann collision integral do not usually fulfill the Einstein's relationship.  When using all three $\eta_D,\kappa_L,\kappa_T$ coefficients independently, the system does not approach equilibrium, as shown in Fig.~\ref{fig:dist_ER}. Although it has been shown that the Langevin equation reproduces the results of the Boltzmann equation under the condition that the scattering angle is small \cite{Aichelin:1984sw}, such condition is often not fulfilled for heavy quark scattering in the pQCD approach and therefore the difference between a Boltzmann dynamics and a reduced Langevin dynamics could be distinct, depending on the the quark mass and regulator, as shown in Ref.~\cite{Das:2013kea}.

We are therefore facing an ambiguity when imposing Einstein's relationship as only two (one) variables among the  $\eta_D,\kappa_L, \kappa_T$ are required in the anisotropic (isotropic) implementation.
Here, we compare the results from different implementation of Einstein's relationship in a static medium, showing in  Fig.~\ref{fig:LBT_vs_LGV}. 
Heavy quarks are initialized with an initial momentum $p_z(0) = 30$ GeV and then propagate in a static medium ($T=0.3$ GeV) for 200 fm/c following different dynamics.
The red dashed line represents the average energy evolution under a linearized Boltzmann dynamics, while the others are the results from Langevin evolution while taking different choices of the drag and momentum transport coefficients.
As shown on the left panel of Fig.~\ref{fig:LBT_vs_LGV}, without involving of the longitudinal momentum transformation $\kappa_L$, heavy quarks lose energy similarly under linearized Boltzmann and Langevin dynamics. 

It should be pointed out that, the coincidence among the implementation of the cases: $\eta_D=ER(\kappa_T), \kappa_T=\kappa_L$, $\kappa_T=\kappa_L=ER(\eta_D)$ is, as a matter of fact, very model-dependent. 
In the leading order pQCD assumption, where the $t$-channel is the main contribution for the energy loss, the relationship between the drag $\eta_D$ and transverse momentum coefficients $\kappa_T$ is approximately close to the Einstein's relationship $\kappa_T \approx 2 E T \eta_D$~\cite{Moore:2004tg}. In this scenario, there is no surprise that the green and orange lines collide with each other.
However, with the consideration of higher-order contribution, mass effects, different choices of regulators, the relationship is not guaranteed.
In an extreme scenario as shown on the right panel of Fig.~\ref{fig:LBT_vs_LGV}, where $\kappa_T$ is quite different from $2 E T \eta_D$, the average energy evolution approaching to equilibrium shows significant deviation between $\eta_D=ER(\kappa_T)$ and $\kappa_T=ER(\eta_D)$ implementations. Such extreme case is achieved here by including only the $s$-channel contribution in the $2\rightarrow 2$ scatterings, which is not intended to describe the reality but used for demonstration purpose. In general, while focusing on the study of average energy loss, the implementation with drag coefficient $\eta_D$ would result in more similarity between Langevin and Boltzmann dynamics.

The average energy is not the only criterion to compare the different implementation. To compare the influence of Einstein's relationship implementation on heavy quark evolution in a realistic medium, we demonstrate the following three cases:
(a) anisotropic case: $(\kappa_L, \kappa_T)$ are known while $\eta_D$ is calculated from Einstein's relationship; 
(b) isotropic case: $\kappa_T$ is known and $\kappa_L = \kappa_T$ and $\eta_D$ is calculated from Einstein's relationship;
(c) isotropic case: $\eta_D$ is known while $\kappa_L=\kappa_T$ are calculated from Einstein's relationship.
Additionally, Einstein's relationship Eqn.~\ref{eqn:ER} holds for the traditional Langevin equation. 
The detailed balance naturally included in the traditional Langevin dynamics breaks down in the improved Langevin equation as a consequence of not including gluon absorption. 
In the improved Langevin implementation, a cut-off for the emitted gluon energy as $\omega = \pi T$ is induced, such that the Boltzmann distribution is still achieved yet with a slightly smaller effective equilibrium temperature~\cite{Cao:2013ita}.

The $R_{\rm AA}$ and $v_2$ with Einstein's relationship imposed are shown in Fig.~\ref{fig:Raa_v2_6fm_ER}.
Just as in the previous subsection, the charm quarks propagate in a (2+1)D hydrodynamical medium and the observables are calculated at the end of the QGP phase. 
Surprisingly, after imposing the Einstein's relationship, the charm quark $R_{\rm AA}$ is significantly smaller compared to the one without imposing Einstein's relationship (Fig.~\ref{fig:Raa_v2_6fm}). 
Since the \Duke coefficients obey the isotropic Einstein's relationship by default, their results are not affected.

When one selects case (b) or (c), the isotropic versions of Einstein's relationship (in those two cases, $\kappa_T$ or $\eta_D$, respectively,  is the only coefficient that controls the interaction strength between charm quarks and the medium), $R_{\rm AA}$ and $v_2$ faithfully reflect the magnitude of $\kappa_T$($\eta_D$). 
A consistently larger \CQPM $\kappa_T$ than \PHSD $\kappa_T$ results in a consistently stronger suppression in $R_{\rm AA}$ and a larger $v_2$, and vice versa.
When one selects case (a) (the anisotropic version of Einstein's relationship), the charm quarks develop the strongest momentum anisotropy, and the peak of $v_2$ has shifted from the lower momentum region $p_T\sim 2.5$ GeV to a higher momentum of around 5 GeV.

Given the ambiguities laid out above, one can question whether the Fokker Planck approach, although very useful to compare different transport approaches, is the right tool for quantitative predictions which can be compared to experimental results. The results for the case that the three transport coefficients are taken as independent - as obtained from the Boltzmann collision integral - and for the case that one imposes the Einstein's relationship differ substantially. In addition the results depend on the arbitrary choice of which of the three transport coefficients is taken over from the Boltzmann collision integral and serves to determine the other two via the Einstein's relationship. These results reinforce the first findings discussed in Ref.~\cite{Das:2013kea} within only a pQCD-like approach.

\section{\label{sec:summary}Conclusion}
The heavy-ion experiments at RHIC and the LHC have provided the community with a rich set of heavy flavor measurements. The main mechanisms driving the strong suppression of high-$p_T$ heavy flavor hadrons and their significant elliptic flow are in general understood, as heavy quarks lose a substantial amount of energy while propagating through the QGP medium. In the low momentum region, the energy loss is dominated by collisional energy loss while in the high momentum region, radiative energy loss plays significant role~\cite{Armesto:2003jh,Armesto:2005iq,Mustafa:2004dr}.
However, the precise determination of the energy loss and the related transport coefficients still lags behind. 
To advance, an improvement of current experimental precision (statistically and systematically) as well as a thorough understanding of the discrepancies observed among theoretical calculations are of crucial importance.

In this report, we have investigated a number of components in the modeling of the heavy quark evolution in heavy-ion collisions in order to evaluate their possible contribution to the determination of the heavy quark transport coefficients in a QGP medium. Key observations we have found include:
\begin{itemize}
\item Charm quarks are sensitive to the history of the QGP evolution and retain information on the entire time evolution from initial condition up to the late stage of the reaction. The calculations confirm that heavy quarks are a very suitable probe to study the QGP properties. 
\item Different initial conditions could cause up to a 20 \%  discrepancy for the final observables $v_2$. This result is obtained using an averaged \trento initial condition and a PHSD initial condition for the same approach for the time evolution.
\item The results for the $v_2$ observable depend on the medium through which the heavy quarks travel. If the expanding plasma is in local equilibrium (hydrodynamics) we obtain -- for the same initial condition -- higher values for $v_2$ as compared to the non-equilibrium PHSD approach. This observation suggest to study whether other observables give additional information on the equilibrium/non-equilibrium expansion of the QGP. In addition, a 15\% of difference in $v_2$ has been shown between heavy quarks propagating in a 2D hydrodynamical medium vs. a 3D hydrodynamical medium. The rapidity distribution is much less dependent on the medium.
\item The inclusion of radiative energy loss has a large effect on the estimation of leading order transport coefficients, particularly to the determination of $\hat{q}$ coefficients (which omit higher order radiative process). In order to make a meaning comparison, one should include all contributions from all processes.
\item The transport coefficient $\kappa_L,  \kappa_T$ and $\eta_D$, calculated from the pQCD cross sections used in the models presented here, do not obey the Einstein's relationship.
Thus the Boltzmann equation, used in these approaches, cannot be consistently reduced to a Langevin equation because the angular distribution of the cross sections cannot be well approximated by retaining only the first two terms of the Taylor expansion. Since for any approach which brings the system asymptotically to a thermal equilibrium the Einstein's relationship has to be fulfilled, one has to make the arbitrary choice which of the three coefficients should be considered as fundamental. The other two are then obtained by the Einstein's relationship. Our results show that the final observables depend strongly on this choice.  
\item Different sources of uncertainties, like different expansion scenarios, different initial conditions and different elementary heavy quark-QGP interactions influence $R_{AA}$ and $v_2$ in a similar way. 
\end{itemize}
To ensure progress in the future, one has to reduce the uncertainties laid out in this manuscript, either by theoretical considerations or by adding new observables into the analysis. Including light hadron observables  may help to limit the variance of the expansion scenarios. Replacing the Langevin approach by a Boltzmann approach helps to eliminate the theoretical uncertainties which are unavoidable if one wants to replace a Boltzmann equation by a Langevin equation, even though care has to be taken regarding heavy quark interactions at small momenta. 

In the future we plan to extend our study to a detailed comparison of the different interaction mechanisms that are implemented in each microscopic model, the effect of the hadronization process, as well as other modeling components such as the dynamics of the pre-equilibrium stage for hydrodynamic models and hadronic final state interaction. 

\section{acknowledgment}
Y.~X. and S.~A.~B. have been supported by the U.S Department of Energy under grant DE-FG02-05ER41367. P.~M, T.~S and E.~B acknowledge the support by the German Academic Exchange Service (DAAD) (T.S., P.M., E.B.); the Deutsche Forschungsgemeinschaft (DFG) under grant No. CRCTR 211 ’Strong-interaction matter under extreme conditions’ (P.M., E.B.); the HGS-HIRe for FAIR (P.M.) and the COST Action THOR, CA15213 (E.B.) J.A and P.G. have been supported by a grant of the region "Pays de la Loire" and by the COST action CA15213 THOR. S.~C is supported by the U.S Department of Energy under grant number DE-SC0013460 and the National Science Fundation under grant number ACI-1550300 (JETSCAPE).

\appendix
\section{$T,p$ dependence of the transport coefficients}
\begin{figure*}
	\includegraphics[width=1.0\textwidth]{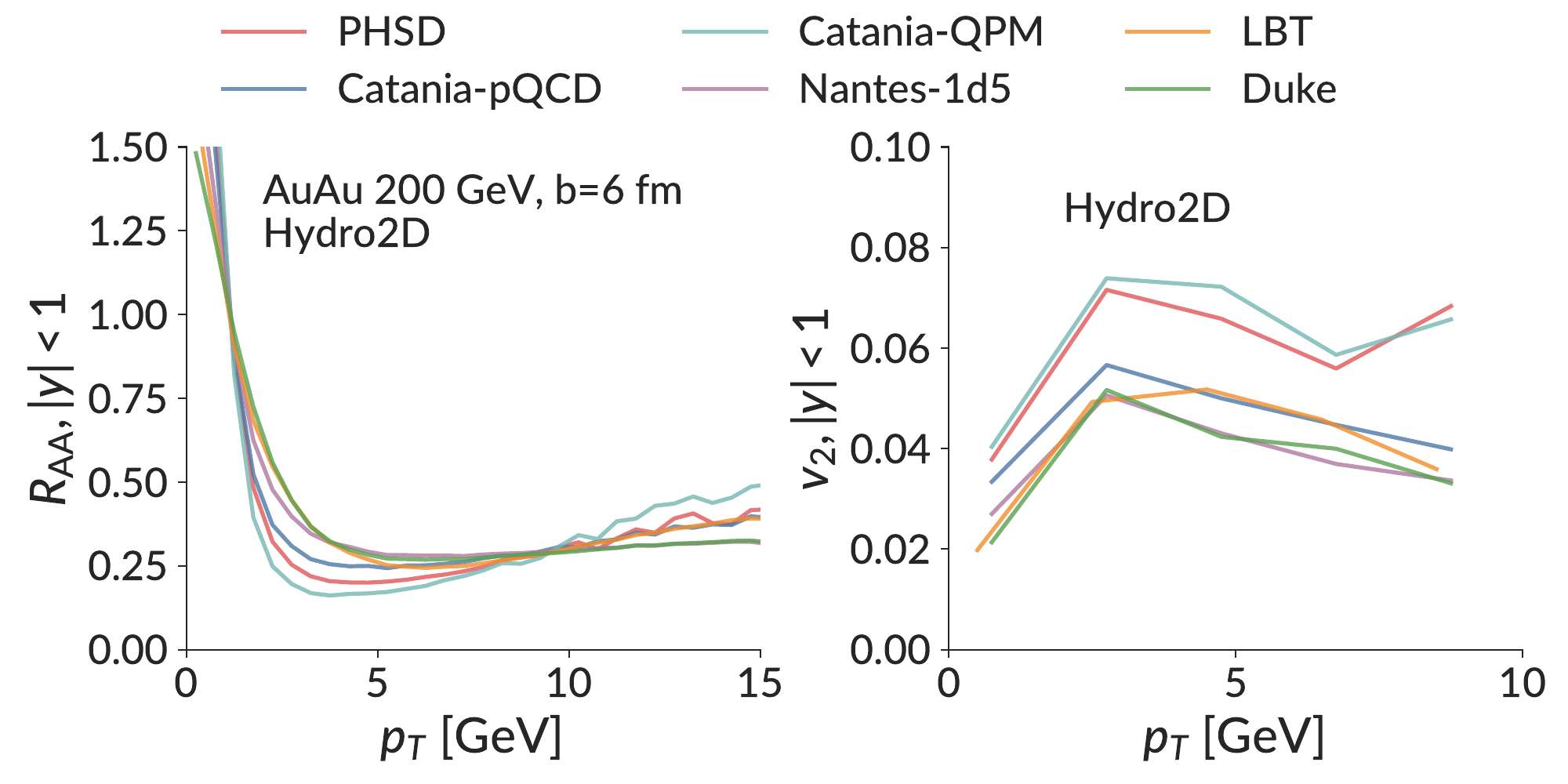}
	\caption{(Color online) Charm quark $R_{\rm AA}$ as a function of $y$ (\textbf{Left}), and $p_T$ (\textbf{Middle}), elliptic flow $v_2$ as a function of $p_T$ (\textbf{Right}) at the end of the QGP phase. The charm quarks are propagating in a hydrodynamical medium simulation for AuAu collisions at 200 GeV with $b=6$ fm/c (\textbf{Upper}), and ``tune2'' assumption for each set of transport coefficients.}
	\label{fig:Raa_v2_6fm_tune2}
\end{figure*}

In this section, we will show that  it is important for each model to describe the $R_{\rm AA}$ and $v_2$ momentum dependence simultaneously in all the momentum regions to which the model can be applied. 
In other words, we will first adjust the coefficients by multiplying each set by a constant $K$ factor, such that we obtain a charm quark $R_{\rm AA}=0.3$ at $p_T=10$ GeV using a Langevin dynamics in a (2+1)D hydrodynamical medium. It is to some extend the analogy to the ``tune2" test in\cite{Cao:2018ews} despite the major difference is that now charm quarks propagate in a dynamical medium instead in a static one.

As shown in Fig.~\ref{fig:Raa_v2_6fm_tune2}, fixing $R_{\rm AA}(p_T=10)=0.3$ makes the variance of the $R_{\rm AA}$ considerably smaller compared to Fig.~\ref{fig:Raa_v2_6fm}, the variance of $v_2$, however, has not improved. This shows the complexity of the dynamics and the independence of both observables.


\begin{thebibliography}{999}
\bibitem{Akiba:2015jwa} 
Y.~Akiba {\it et al.},
arXiv:1502.02730 [nucl-ex].

\bibitem{Andronic:2015wma} 
A.~Andronic {\it et al.},
Eur.\ Phys.\ J.\ C {\bf 76}, no. 3, 107 (2016)
doi:10.1140/epjc/s10052-015-3819-5
[arXiv:1506.03981 [nucl-ex]].


\bibitem{Rapp:2009my} 
R.~Rapp and H.~van Hees,
doi:10.1142/9789814293297-0003
arXiv:0903.1096 [hep-ph].

\bibitem{Zakharov:2000iz} 
B.~G.~Zakharov,
JETP Lett.\  {\bf 73}, 49 (2001)
[Pisma Zh.\ Eksp.\ Teor.\ Fiz.\  {\bf 73}, 55 (2001)]
doi:10.1134/1.1358417
[hep-ph/0012360].

\bibitem{Gyulassy:2002yv} 
M.~Gyulassy, P.~Levai and I.~Vitev,
Phys.\ Rev.\ D {\bf 66}, 014005 (2002)
doi:10.1103/PhysRevD.66.014005
[nucl-th/0201078].

\bibitem{Baier:1996sk} 
R.~Baier, Y.~L.~Dokshitzer, A.~H.~Mueller, S.~Peigne and D.~Schiff,
Nucl.\ Phys.\ B {\bf 484}, 265 (1997)
doi:10.1016/S0550-3213(96)00581-0
[hep-ph/9608322].

\bibitem{Wang:2001ifa} 
X.~N.~Wang and X.~f.~Guo,
Nucl.\ Phys.\ A {\bf 696}, 788 (2001)
doi:10.1016/S0375-9474(01)01130-7
[hep-ph/0102230].

\bibitem{Banerjee:2011ra} 
D.~Banerjee, S.~Datta, R.~Gavai and P.~Majumdar,
Phys.\ Rev.\ D {\bf 85}, 014510 (2012)
doi:10.1103/PhysRevD.85.014510
[arXiv:1109.5738 [hep-lat]].


\bibitem{Prino:2016cni} 
F.~Prino and R.~Rapp,
J.\ Phys.\ G {\bf 43}, no. 9, 093002 (2016)
doi:10.1088/0954-3899/43/9/093002
[arXiv:1603.00529 [nucl-ex]].

\bibitem{Chesler:2013urd} 
P.~M.~Chesler, M.~Lekaveckas and K.~Rajagopal,
JHEP {\bf 1310}, 013 (2013)
doi:10.1007/JHEP10(2013)013
[arXiv:1306.0564 [hep-ph]].

\bibitem{Braaten:2002yt} 
E.~Braaten, Y.~Jia and T.~Mehen,
Phys.\ Rev.\ Lett.\  {\bf 89}, 122002 (2002)
doi:10.1103/PhysRevLett.89.122002
[hep-ph/0205149].

\bibitem{Lin:2003jy} 
Z.~w.~Lin and D.~Molnar,
Phys.\ Rev.\ C {\bf 68}, 044901 (2003)
doi:10.1103/PhysRevC.68.044901
[nucl-th/0304045].


\bibitem{Laine:2011is} 
M.~Laine,
JHEP {\bf 1104}, 124 (2011)
doi:10.1007/JHEP04(2011)124
[arXiv:1103.0372 [hep-ph]].

\bibitem{He:2011yi} 
M.~He, R.~J.~Fries and R.~Rapp,
Phys.\ Lett.\ B {\bf 701}, 445 (2011)
doi:10.1016/j.physletb.2011.06.019
[arXiv:1103.6279 [nucl-th]].


\bibitem{Cao:2018ews} 
  S.~Cao {\it et al.},
  arXiv:1809.07894 [nucl-th].
  
\bibitem{Rapp:2018qla} 
R.~Rapp {\it et al.},
arXiv:1803.03824 [nucl-th].



\bibitem{Song:2015sfa} 
T.~Song, H.~Berrehrah, D.~Cabrera, J.~M.~Torres-Rincon, L.~Tolos, W.~Cassing and E.~Bratkovskaya,
Phys.\ Rev.\ C {\bf 92}, no. 1, 014910 (2015)
doi:10.1103/PhysRevC.92.014910
[arXiv:1503.03039 [nucl-th]].

\bibitem{Song:2007fn} 
H.~Song and U.~W.~Heinz,
Phys.\ Lett.\ B {\bf 658}, 279 (2008)
doi:10.1016/j.physletb.2007.11.019
[arXiv:0709.0742 [nucl-th]].

\bibitem{Plumari:2012ep} 
S.~Plumari, A.~Puglisi, F.~Scardina and V.~Greco,
Phys.\ Rev.\ C {\bf 86}, 054902 (2012)
doi:10.1103/PhysRevC.86.054902
[arXiv:1208.0481 [nucl-th]].

\bibitem{Das:2015ana} 
S.~K.~Das, F.~Scardina, S.~Plumari and V.~Greco,
Phys.\ Lett.\ B {\bf 747}, 260 (2015)
doi:10.1016/j.physletb.2015.06.003
[arXiv:1502.03757 [nucl-th]].

\bibitem{Plumari:2011mk} 
S.~Plumari, W.~M.~Alberico, V.~Greco and C.~Ratti,
Phys.\ Rev.\ D {\bf 84}, 094004 (2011)
doi:10.1103/PhysRevD.84.094004
[arXiv:1103.5611 [hep-ph]].


\bibitem{Scardina:2017ipo} 
F.~Scardina, S.~K.~Das, V.~Minissale, S.~Plumari and V.~Greco,
Phys.\ Rev.\ C {\bf 96}, no. 4, 044905 (2017)
doi:10.1103/PhysRevC.96.044905
[arXiv:1707.05452 [nucl-th]].

\bibitem{Gossiaux:2008jv} 
P.~B.~Gossiaux and J.~Aichelin,
Phys.\ Rev.\ C {\bf 78}, 014904 (2008)
doi:10.1103/PhysRevC.78.014904
[arXiv:0802.2525 [hep-ph]].


\bibitem{Nahrgang:2013saa} 
M.~Nahrgang, J.~Aichelin, P.~B.~Gossiaux and K.~Werner,
Phys.\ Rev.\ C {\bf 90}, no. 2, 024907 (2014)
doi:10.1103/PhysRevC.90.024907
[arXiv:1305.3823 [hep-ph]].

\bibitem{Cao:2016gvr} 
S.~Cao, T.~Luo, G.~Y.~Qin and X.~N.~Wang,
Phys.\ Rev.\ C {\bf 94}, no. 1, 014909 (2016)
doi:10.1103/PhysRevC.94.014909
[arXiv:1605.06447 [nucl-th]].

\bibitem{Cao:2017hhk} 
S.~Cao, T.~Luo, G.~Y.~Qin and X.~N.~Wang,
Phys.\ Lett.\ B {\bf 777}, 255 (2018)
doi:10.1016/j.physletb.2017.12.023
[arXiv:1703.00822 [nucl-th]].


\bibitem{Cao:2013ita} 
S.~Cao, G.~Y.~Qin and S.~A.~Bass,
Phys.\ Rev.\ C {\bf 88}, 044907 (2013)
doi:10.1103/PhysRevC.88.044907
[arXiv:1308.0617 [nucl-th]].

\bibitem{Cao:2015hia} 
S.~Cao, G.~Y.~Qin and S.~A.~Bass,
Phys.\ Rev.\ C {\bf 92}, no. 2, 024907 (2015)
doi:10.1103/PhysRevC.92.024907
[arXiv:1505.01413 [nucl-th]].

\bibitem{Xu:2017hgt} 
Y.~Xu, M.~Nahrgang, J.~E.~Bernhard, S.~Cao and S.~A.~Bass,
Nucl.\ Phys.\ A {\bf 967}, 668 (2017)
doi:10.1016/j.nuclphysa.2017.05.035
[arXiv:1704.07800 [nucl-th]].


\bibitem{He:2013zua} 
M.~He, H.~van Hees, P.~B.~Gossiaux, R.~J.~Fries and R.~Rapp,
Phys.\ Rev.\ E {\bf 88}, 032138 (2013)
doi:10.1103/PhysRevE.88.032138
[arXiv:1305.1425 [nucl-th]].

\bibitem{vanHees:2004gq} 
H.~van Hees and R.~Rapp,
Phys.\ Rev.\ C {\bf 71}, 034907 (2005)
doi:10.1103/PhysRevC.71.034907
[nucl-th/0412015].



\bibitem{Moore:2004tg} 
G.~D.~Moore and D.~Teaney,
Phys.\ Rev.\ C {\bf 71}, 064904 (2005)
doi:10.1103/PhysRevC.71.064904
[hep-ph/0412346].


\bibitem{Ding:2012sp} 
H.~T.~Ding, A.~Francis, O.~Kaczmarek, F.~Karsch, H.~Satz and W.~Soeldner,
Phys.\ Rev.\ D {\bf 86}, 014509 (2012)
doi:10.1103/PhysRevD.86.014509
[arXiv:1204.4945 [hep-lat]].


\bibitem{Borsanyi:2010bp} 
S.~Borsanyi {\it et al.} [Wuppertal-Budapest Collaboration],
JHEP {\bf 1009}, 073 (2010)
doi:10.1007/JHEP09(2010)073
[arXiv:1005.3508 [hep-lat]].

\bibitem{Bhattacharya:2014ara} 
T.~Bhattacharya {\it et al.},
Phys.\ Rev.\ Lett.\  {\bf 113}, no. 8, 082001 (2014)
doi:10.1103/PhysRevLett.113.082001
[arXiv:1402.5175 [hep-lat]].







\bibitem{Cacciari:1998it} 
M.~Cacciari, M.~Greco and P.~Nason,
JHEP {\bf 9805}, 007 (1998)
doi:10.1088/1126-6708/1998/05/007
[hep-ph/9803400].

\bibitem{Cacciari:2012ny} 
M.~Cacciari, S.~Frixione, N.~Houdeau, M.~L.~Mangano, P.~Nason and G.~Ridolfi,
JHEP {\bf 1210}, 137 (2012)
doi:10.1007/JHEP10(2012)137
[arXiv:1205.6344 [hep-ph]].



\bibitem{Eskola:2009uj} 
K.~J.~Eskola, H.~Paukkunen and C.~A.~Salgado,
JHEP {\bf 0904}, 065 (2009)
doi:10.1088/1126-6708/2009/04/065
[arXiv:0902.4154 [hep-ph]].

\bibitem{Moreland:2014oya} 
  J.~S.~Moreland, J.~E.~Bernhard and S.~A.~Bass,
  Phys.\ Rev.\ C {\bf 92}, no. 1, 011901 (2015)
  doi:10.1103/PhysRevC.92.011901
  [arXiv:1412.4708 [nucl-th]].


\bibitem{Schenke:2013dpa} 
B.~Schenke, P.~Tribedy and R.~Venugopalan,
Phys.\ Rev.\ C {\bf 89}, no. 2, 024901 (2014)
doi:10.1103/PhysRevC.89.024901
[arXiv:1311.3636 [hep-ph]].

\bibitem{Karpenko:2013wva} 
  I.~Karpenko, P.~Huovinen and M.~Bleicher,
  Comput.\ Phys.\ Commun.\  {\bf 185}, 3016 (2014)
  doi:10.1016/j.cpc.2014.07.010
  [arXiv:1312.4160 [nucl-th]].

\bibitem{Xu:2017pna} 
Y.~Xu, P.~Moreau, T.~Song, M.~Nahrgang, S.~A.~Bass and E.~Bratkovskaya,
Phys.\ Rev.\ C {\bf 96}, no. 2, 024902 (2017)
doi:10.1103/PhysRevC.96.024902
[arXiv:1703.09178 [nucl-th]].


\bibitem{Qiu:2011hf} 
Z.~Qiu, C.~Shen and U.~Heinz,
Phys.\ Lett.\ B {\bf 707}, 151 (2012)
doi:10.1016/j.physletb.2011.12.041
[arXiv:1110.3033 [nucl-th]].

\bibitem{Uphoff:2012gb} 
J.~Uphoff, O.~Fochler, Z.~Xu and C.~Greiner,
Phys.\ Lett.\ B {\bf 717}, 430 (2012)
doi:10.1016/j.physletb.2012.09.069
[arXiv:1205.4945 [hep-ph]].


\bibitem{Gossiaux:2011ea} 
P.~B.~Gossiaux, S.~Vogel, H.~van Hees, J.~Aichelin, R.~Rapp, M.~He and M.~Bluhm,
[arXiv:1102.1114 [hep-ph]].

\bibitem{Alberico:2011zy} 
W.~M.~Alberico, A.~Beraudo, A.~De Pace, A.~Molinari, M.~Monteno, M.~Nardi and F.~Prino,
Eur.\ Phys.\ J.\ C {\bf 71}, 1666 (2011)
doi:10.1140/epjc/s10052-011-1666-6
[arXiv:1101.6008 [hep-ph]].

\bibitem{Berrehrah:2014tva}
  H.~Berrehrah, P.~B.~Gossiaux, J.~Aichelin, W.~Cassing, J.~M.~Torres-Rincon and E.~Bratkovskaya,
  Phys.\ Rev.\ C {\bf 90} (2014) 051901
  doi:10.1103/PhysRevC.90.051901
  [arXiv:1406.5322 [hep-ph]].

  
\bibitem{Gossiaux:2006yu} 
P.~B.~Gossiaux, V.~Guiho and J.~Aichelin,
J.\ Phys.\ G {\bf 32}, S359 (2006).
doi:10.1088/0954-3899/32/12/S44


\bibitem{vanHees:2005wb} 
H.~van Hees, V.~Greco and R.~Rapp,
Phys.\ Rev.\ C {\bf 73}, 034913 (2006)
doi:10.1103/PhysRevC.73.034913
[nucl-th/0508055].

\bibitem{Das:2010tj} 
S.~K.~Das, J.~e.~Alam and P.~Mohanty,
Phys.\ Rev.\ C {\bf 82}, 014908 (2010)
doi:10.1103/PhysRevC.82.014908
[arXiv:1003.5508 [nucl-th]].


\bibitem{Aichelin:1984sw}
  J.~Aichelin,
  Nucl.\ Phys.\ A {\bf 411} (1983) 474.
  doi:10.1016/0375-9474(83)90541-9


\bibitem{Das:2013kea} 
  S.~K.~Das, F.~Scardina, S.~Plumari and V.~Greco,
  Phys.\ Rev.\ C {\bf 90}, 044901 (2014)
  doi:10.1103/PhysRevC.90.044901
  [arXiv:1312.6857 [nucl-th]].
  

  



\bibitem{Armesto:2003jh} 
N.~Armesto, C.~A.~Salgado and U.~A.~Wiedemann,
Phys.\ Rev.\ D {\bf 69}, 114003 (2004)
doi:10.1103/PhysRevD.69.114003
[hep-ph/0312106].

\bibitem{Armesto:2005iq} 
N.~Armesto, A.~Dainese, C.~A.~Salgado and U.~A.~Wiedemann,
Phys.\ Rev.\ D {\bf 71}, 054027 (2005)
doi:10.1103/PhysRevD.71.054027
[hep-ph/0501225].

\bibitem{Mustafa:2004dr} 
M.~G.~Mustafa,
Phys.\ Rev.\ C {\bf 72}, 014905 (2005)
doi:10.1103/PhysRevC.72.014905
[hep-ph/0412402].







\bibitem{Gossiaux:2009qf} 
P.~B.~Gossiaux and J.~Aichelin,
Nucl.\ Phys.\ A {\bf 830}, 203C (2009)
doi:10.1016/j.nuclphysa.2009.10.015
[arXiv:0907.4329 [hep-ph]].





\bibitem{Francis:2015daa} 
A.~Francis, O.~Kaczmarek, M.~Laine, T.~Neuhaus and H.~Ohno,
Phys.\ Rev.\ D {\bf 92}, no. 11, 116003 (2015)
doi:10.1103/PhysRevD.92.116003
[arXiv:1508.04543 [hep-lat]].


\end{thebibliography}
\end{document}